\documentclass[aoas]{imsart}

\usepackage{centernot}
\usepackage{bm}
\usepackage{xcolor}
\usepackage{booktabs}
\usepackage{float}
\usepackage{dsfont}
\usepackage{graphicx}
\usepackage{tikz}
\usepackage{url}
\usepackage{threeparttable}
\usetikzlibrary{spy}
\usetikzlibrary{backgrounds}
\usetikzlibrary{shapes,decorations,arrows,calc,arrows.meta,fit,positioning}
\RequirePackage{amsthm,amsmath,amsfonts,amssymb}
\RequirePackage[authoryear]{natbib}

\startlocaldefs
\theoremstyle{plain}
\theoremstyle{remark}
\newtheorem{assumption}{Assumption}

 \newcommand{\ind}{\perp\!\!\!\!\perp}
 \newcommand{\notind}{\centernot{\ind}}
 
\endlocaldefs

\begin{document}

\begin{frontmatter}
\title{Joint mixed-effects models for causal inference in clustered network-based observational studies\protect\thanks{VM is supported by doctoral fellowships from Natural Sciences and Engineering Research Council of Canada (NSERC) and the Fonds de Recherche du Québec (FRQ) - Nature et Technologie. EEMM acknowledges support from an NSERC Discovery Grant. EEMM is a Canada Research Chair (Tier 1) in Statistical Methods for Precision Medicine and acknowledges the support of a Chercheur-boursier de mérite career award from the FRQ - Santé.}}
\runtitle{Joint mixed-effects models for causal inference with interference}

\begin{aug}
\author[A]{\fnms{Vanessa}~\snm{McNealis}\ead[label=e1]{vanessa.mcnealis@mail.mcgill.ca}},
\author[B]{\fnms{Erica E. M.}~\snm{Moodie}\ead[label=e2]{erica.moodie@mcgill.ca}}
\and
\author[C]{\fnms{Nema}~\snm{Dean}\ead[label=e3]{nema.dean@glasgow.ac.eu}}
\address[A]{Department of Epidemiology and Biostatistics, McGill University\printead[presep={,\ }]{e1}}
\address[B]{Department of Epidemiology and Biostatistics, McGill University\printead[presep={,\ }]{e2}}

\address[C]{School of Mathematics and Statistics, University of Glasgow \printead[presep={,\ }]{e3}}
\end{aug}

\begin{abstract}
Causal inference on populations embedded in social networks poses technical challenges, since the typical no interference assumption frequently does not hold. Existing methods developed in the context of network interference rely upon the assumption of no unmeasured confounding. However, when faced with multilevel network data, there may be a latent factor influencing both the exposure and the outcome at the cluster level. We propose a Bayesian inference approach that combines a joint mixed-effects model for the outcome and the exposure with direct standardization to identify and estimate causal effects in the presence of network interference and unmeasured cluster confounding. In simulations, we compare our proposed method with linear mixed– and fixed effects models and show that unbiased estimation is achieved using the joint model. Having derived valid tools for estimation, we examine the effect of maternal college education on adolescent school performance using data from the National Longitudinal Study of Adolescent Health.
\end{abstract}


\end{frontmatter}


\section{Introduction}
Much of the theoretical development in causal inference has been within the counterfactual framework, which typically employs a “no interference” assumption, in that the potential outcome of an individual is not affected by the treatments of another. In many sociological settings, the “no interference” assumption is implausible, and traditional methods will fail whenever social interaction influences the mechanisms relating a treatment to an outcome. In such cases, the treatment or exposure assigned to an individual will also affect their close neighbors' outcomes, yielding \textit{spillover} effects. A highly relevant example is in studies of adolescents, where the influence of peers takes on a much greater role than in early childhood or later life. Consider the impact of maternal college education on adolescent school performance: it is entirely plausible that an adolescent's performance is influenced both by their home environment and that of their peers.

 Peer influence has long been of interest to social scientists, since it can potentially inform the optimal organization of schools, neighborhoods, workplaces, and other structures whereby individuals interact. In educational research, there has been mounting evidence suggesting that classroom composition, notably in terms of ability or achievement distribution, can have an influence on learning outcomes \citep{hattie2002classroom, jones2016does}. For instance, using administrative public school data gathered by the Texas Education Agency, \cite{hoxby2000peer} found that a change in one point in peers' reading scores is associated with an increase in a student's own score between 0.15 and 0.40. Leveraging data from the National Longitudinal Study of Adolescent Health (Add
Health), \cite{bifulco2011effect} found that having a higher proportion of classmates with college-educated mothers was associated with a lower risk of dropping out and a higher likelihood of attending college. \cite{fletcher2020consequences} made use of the friendship nominations data in Add Health and found that a greater proportion of friends with college-educated mothers is associated with a higher Grade Point Average (GPA), with evidence of a differential effect by sex. This body of evidence supports the idea of spillover effects of home environment on academic performance.

Social network data are often collected when such interference is at play, and many of the settings in which spillover effects are being estimated are multilevel (e.g., social networks of students nested within schools). The Add Health study is one example of such multilevel structure. When faced with multilevel network data, such as one might find in schools or geographical communities, for example, there may be an unmeasured, latent factor influencing both the exposure and the outcome at the cluster level. This tendency for peers, such as school mates or neighborhood residents, to share some unobserved contextual factors is termed \textit{contextual confounding} \citep{vanderweele2013social}. For spatially structured data, this problem can arise when the statistical model fails to account for an important environmental determinant that in itself is spatially structured and thus causes spatial structuring in the response \citep{f2007methods}. In the context of adolescent school performance, there could be a latent trait influencing both the average maternal education as well as the adolescents' overall performance at the school level. If it is unmeasured, such latent trait could be geographical location for instance, since it may encompass area level violence, pollution, or access to quiet learning spaces. The presence of an unmeasured confounder can invalidate causal inferences drawn in observational network-based studies by distorting the effect that we observe.

Causal inference is inherently a missing data problem and the problem of unmeasured contextual confounding can be conceived as a nonignorable missingness mechanism \citep{ding2018causal}. Nonignorable missingness or treatment assignment mechanisms are typically addressed by imposing additional parametric assumptions. In the frequentist paradigm, joint mixed-effects models were originally proposed to handle nonignorable missing data when estimating the parameters of an outcome data-generating model \citep{follmann1995approximate, little2019statistical}. \cite{shardell2018joint} combined parametric joint mixed-effects models for a study outcome and an exposure with g-computation to identify and estimate causal effects in longitudinal settings under unmeasured confounding. In the Bayesian paradigm, \cite{papadogeorgou2023spatial} proposed an approach based on simultaneous modeling of the exposure and outcome to account for the presence of spatially-structured unmeasured confounders. \cite{nobre2023impact} used a Bayesian multilevel approach to study the effect of directly observed therapy on the rate of cure of Tuberculosis while mitigating the impact of unobserved contextual factors shared by individuals within different municipalities across the state of São Paulo, Brazil. \cite{xu2023causal} proposed Bayesian multivariate generalized linear mixed-effects models to assess dynamic treatment regimes while accounting for unmeasured patient heterogeneity.

Existing methods that accommodate multilevel networks, be it inverse probability-of-treatment weighting (IPW), outcome regression, or doubly robust estimation, predominantly rely upon the assumption of no unmeasured confounding \citep{lee2021estimating,forastiere2021identification, mcnealis2023doubly}. In this work, we address an important gap by relaxing the unconfoundedness assumption and providing a Bayesian framework for causal inference in the multilevel network setting, in which heterogeneity is due to clustering of individuals in different subgraphs. We propose an approach based on a joint mixed-effects model (JMM) combined with Bayesian standardization to estimate direct and indirect effects of a binary treatment on a continuous outcome in the presence of unmeasured contextual confounding. We study frequentist properties of the JMM approach in simulation studies, which show that valid inference can be obtained in comparison with more naive approaches that either assume ignorability of the treatment assignment (linear mixed-effects model) or ignore the clustered nature of the data altogether (fixed effects model).

The paper is structured as follows. In Section \ref{section:addhealth}, we introduce the Add Health study and our motivating question examining the direct and indirect effect of maternal education on school performance. 
In Sections \ref{section:notation} and \ref{section:estimands}, we introduce the counterfactual framework and causal assumptions under the network interference setting and the relevant causal estimands for this setting. In Section \ref{section:inferentialmethod}, we describe the proposed JMM approach along with the Bayesian standardization procedure used to perform posterior inference for causal effects. In Section \ref{section:simulations}, we assess frequentist properties of the proposed estimator in simulation studies. The method is then utilized to assess the direct and spillover effects of maternal college education on adolescent school performance using the Add Health network data in Section \ref{section:application}. We close with a general discussion in Section \ref{section:discussion}.

\section{The Add Health study}
\label{section:addhealth}
Add Health is a longitudinal study of a nationally representative sample of 90,118 adolescent students in grades 7-12 in the
United States in 1994-95 who were followed throughout adolescence \citep{harris2013add}. The original purpose of the study was to 
understand the determinants of adolescent health and health behaviour. Add Health used a school-based design that selected 80 high schools and associated feeder schools (typically a middle school) using a sampling frame derived from the Quality Education Database. The final sample comprised 132 schools with communities located in urban, suburban, and rural areas of the country.

\begin{figure}[ht]
    \centering
        \includegraphics[scale=0.3]{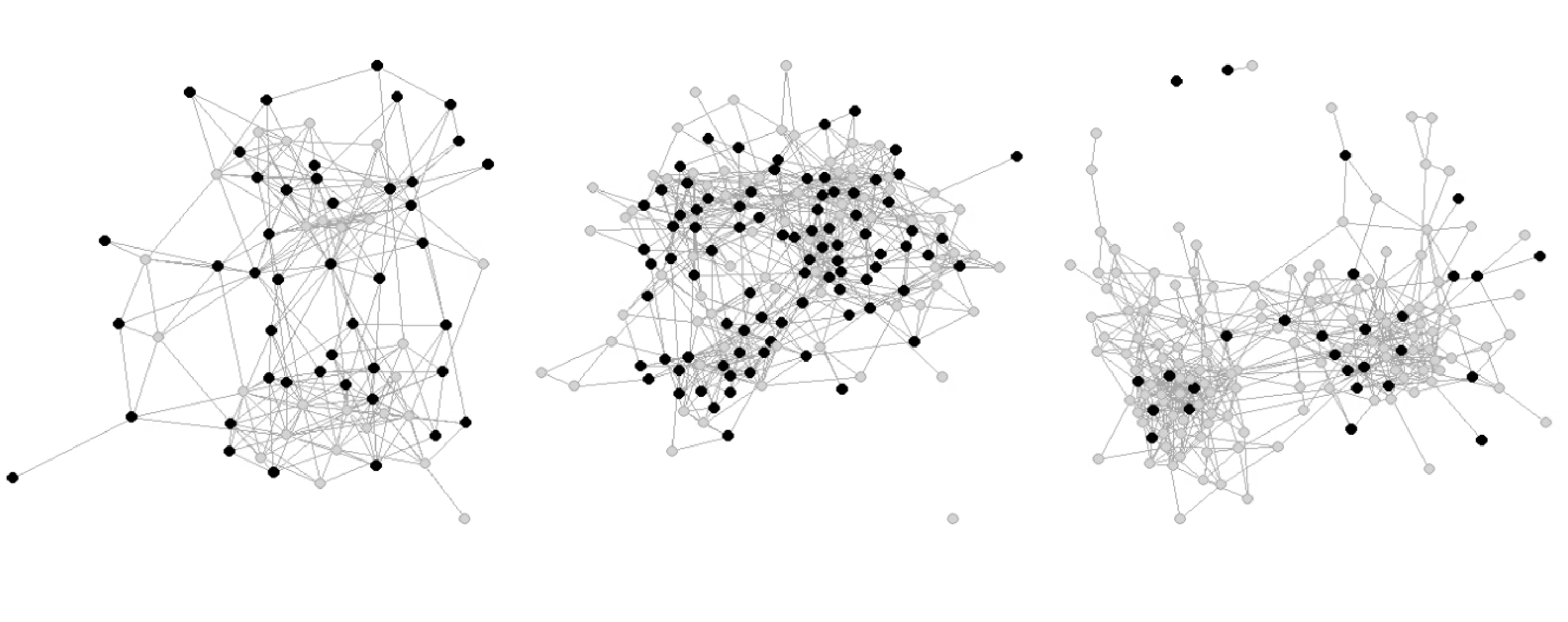}
    \caption[Network visualizations formed by three schools in Add Health.]{Three schools in the Add Health network of 71, 159 and 160 students, respectively, where an edge between two students indicates that either one nominated the other as a friend and black-coloured nodes represent students with college-educated mothers. Note that ties that exist across different schools are not shown in this visualization. }
    \label{fig:threeschools}
\end{figure}

In addition to socio-demographic and academic achievement characteristics, peer network data were obtained in the in-school questionnaire, in which adolescents nominated up to five best male and five best female friends from the school roster. We mapped an undirected network to friendship nomination data from the first wave of data collection in 1994-95 of Add Health. Following a reciprocal definition of friendship, we drew an edge between students $i$ and $j$ if student $i$ listed $j$ as a friend in the in-school survey, or $j$ listed $i$ as a friend, or both. For the purpose of this analysis, we selected a random sample of 16 schools. Figure \ref{fig:threeschools} represents the three smallest schools in the subsample, with students with a college-educated mother represented by black-colored nodes. The network characteristics and distribution of attributes are summarized in Tables \ref{tab:table1} and \ref{tab:table12}, respectively, which can be found in Appendix \ref{section:appendixa}. The Add Health subnetwork has average degree 6.65 (SD=3.88) and density 0.0008. The transitivity, which measures the extent to which nodes tend to cluster together, is 0.171. Perfect transitivity implies that if two nodes $x$ and $z$ are connected to a another node $y$, then systematically $x$ and $z$ are also connected. The assortativity by maternal education, which is 0.123 globally, is the extent to which nodes with matching values of maternal education tend to form ties in the network.  

Previous work has examined the causal effect of peers on academic success under a counterfactual framework, focusing on the indirect (spillover) effects of having a college-educated mother on GPA \citep{mcnealis2023doubly}. As individual-level data are available for each school in Add Health, it is expected that students attend the same school and thus residing in the same area share latent traits affecting both their propensity to have a college-educated mother and their overall grades. Thus, we suspect unmeasured subgraph-level confounders influencing both the overall socio-economic status and the overall academic performance at the school level. This raises the question of whether subgraph-level random effects should be included when modelling the quantities of interest. Therefore, we follow the Bayesian paradigm and discuss a procedure that jointly models propensity score and the outcome to estimate direct and indirect effects of maternal education on academic performance. Throughout the text, we will refer to maternal education as a (self-selected) treatment. 

\section{Notation and causal assumptions}
\label{section:notation}
Consider an undirected network $G=(\mathcal{N}, \mathbb{E})$, where $\mathcal{N} = \{1, 2, \ldots, N\}$ is a set of $N$ nodes and $\mathbb{E}$ is a set of edges $\mathbb{E}$ with generic element $\{i,j\} = \{j,i\}$ denoting the presence of an edge between nodes $i$ and $j$. The sociomatrix $\bm{A}$ associated to $G$ has an entry $A_{ij} = 1$ if $\{i, j\}$ is in $\mathbb{E}$, and 0 otherwise. We assume that the set of nodes in this network is clustered, in that it can be written as an ensemble of $m$ disjoint subsets $\mathcal{N}_{\nu}$, $\nu = 1, \ldots, m$, each of order $|\mathcal{N}_{\nu}|=N_{\nu}$. The $m$ subsets could for instance correspond to different schools or classrooms. We define $G_{\nu} = (\mathcal{N}_{\nu}, \mathbb{E}_{\nu})$ to be the induced subgraph associated with the set of nodes $\mathcal{N}_{\nu}$. Note that this definition does not preclude the existence of ties across the different node sets $\mathcal{N}_{\nu}$. We define the neighborhood $\mathcal{N}_{\nu i}$ of node $i$ in cluster $\nu$, $i=1,2,\ldots, N_{\nu}$, as the set of nodes $j$ for which $\{i,j\} \in \mathbb{E}_{\nu}$ and the degree of node $i$ is denoted $d_{\nu i} = |\mathcal{N}_{\nu i}|$. Since the subgraphs $\{G_{\nu}, \nu = 1, \ldots, m\}$ are not necessarily disjoint, nodes in the immediate neighborhood $\mathcal{N}_{\nu i}$ may belong to a different subset than $\mathcal{N}_{\nu}$, say $\mathcal{N}_{\upsilon}$, $\upsilon \neq \nu$. Additionally, we denote $\mathcal{N}_{\nu i}^* = \mathcal{N}_{\nu i} \cup {i}$ as the union of the node $i$ and the nodes in $\mathcal{N}_{\nu i}$. It follows that to each node $i$ is associated a partition of $\mathcal{N}$, that is $\{i, \mathcal{N}_{\nu i}, \mathcal{N}_{\nu -i}\}$, where  $\mathcal{N}_{\nu -i} = \mathcal{N}\backslash \mathcal{N}_{\nu i}^*$. 
 
 A node $i$ in cluster or subgraph $\nu$ has an observed outcome $Y_{\nu i}$, an assigned binary treatment $Z_{\nu i}$ and a vector of covariates $\bm{X}_{\nu i}$. Covariates may include an intercept, unit $i$'s individual covariates, or covariates corresponding to unit $i$'s neighborhood or subgraph. Let 
 $\bm{Z}_{\mathcal{N}_{\nu i}} = (Z_{j_1}, \ldots, Z_{j_{d_{\nu i}}})'$ denote the observed neighborhood treatment vector for node $i$ in subgraph $\nu$, where $j_{k} \in \mathcal{N}_{\nu i}, k = 1, \ldots, d_{\nu i}$. 
We extend typical causal assumptions to include the presence of network interference. Let $y_{\nu i}(z_{\nu i},\bm{z}_{\mathcal{N}_{\nu i}}, \bm{z}_{\mathcal{N}_{\nu -i}})$ denote the potential outcome for individual $i$ in subgraph $\nu$ if they (and their neighbors and non-neighbors) receive the treatment assignment $(z_{\nu i},\bm{z}_{\mathcal{N}_{\nu i}}, \bm{z}_{\mathcal{N}_{\nu -i}})$. While in principle this allows the potential outcome to depend on the entire treatment assignment in the network, we will soon restrict the number of potential outcomes using the known individual connections and assuming an exposure mapping; see Assumption \ref{assum:stratified} below \citep{aronow2017estimating, forastiere2021identification}. We first make the consistency assumption, which states that there cannot be multiple versions of a treatment.
\begin{assumption}[Consistency]
\label{assump:consist}
For $\nu \in \{1, \ldots, m \}$, $i \in \mathcal{N}_{\nu}$, $Y_{\nu i} = y_{\nu i}(Z_{\nu i}, \bm{Z}_{\mathcal{N}_{\nu i}}, \bm{Z}_{\mathcal{N}_{\nu -i}})$.
\end{assumption}
\begin{assumption}[Stratified interference]
\label{assum:stratified}
For $\nu \in \{1, \ldots, m \}$, $i \in \mathcal{N}_{\nu}$, $\forall \ \bm{z}_{\mathcal{N}_{\nu -i}}, \bm{z}_{\mathcal{N}_{\nu -i}}'$, and $\forall$ $\bm{z}_{\mathcal{N}_{\nu i}}, \bm{z}_{\mathcal{N}_{\nu i}}'$ such that $\phi(\bm{z}_{\mathcal{N}_{\nu i}}) = \phi(\bm{z}_{\mathcal{N}_{\nu i}}')$ given a function $\phi: \{0,1\}^{d_\nu i} \rightarrow \Phi_{\nu i}$, then, $$y_{\nu i}(z_{\nu i}, \bm{z}_{\mathcal{N}_{\nu i}}, \bm{z}_{\mathcal{N}_{\nu -i}})  =  y_{\nu i}(z_{\nu i}, \bm{z}_{\mathcal{N}_{\nu i}}, \bm{z}_{\mathcal{N}_{\nu -i}}').$$
\end{assumption}
 For simplicity, in what follows we assume that the function $\phi$ corresponds to the total exposure value among the neighbors, that is, $\phi: \{0,1\}^{d_{\nu i}} \rightarrow \Phi_{\nu i}$ defined by $\phi(\bm{z}_{\mathcal{N}_{\nu i}}) = \sum_{j \in \mathcal{N}_{\nu i}} z_{\nu j}$, where $\Phi_{\nu i} \in \{0, 1, \ldots, d_{\nu i}\}$. Because of Assumption 2, we can drop the dependence of $y_{\nu i}(\cdot)$ on the nodes in $\mathcal{N}_{\nu -i}$ and write more simply $y_{\nu i}(z_{\nu i}, z_{\mathcal{N}_{\nu i}})$. 
 
 In the presence of unmeasured contextual confounding, the covariates $\bm{X}_{\nu i}$ are no longer a sufficient conditioning set for unconfoundedness of the treatment assignment, i.e.,  $y_{\nu i}(z_{\nu i}, \bm{z}_{\mathcal{N}_{\nu i}}) \notind Z_{\nu i}, \bm{Z}_{\mathcal{N}_{\nu i}} | \bm{X}_{\nu i}.$ 
 We assume however that exchangeability holds conditional on measured covariates, and the subgraph-level value of an unmeasured covariate.
 \begin{assumption}[Latent ignorability] 
 \label{assump:condexch} There exists unmeasured subgraph-level covariates $\bm{b}^y = (\bm{b}_{1}^y, \bm{b}_{2}^y, \ldots, \bm{b}_{m}^y)$, each of dimension $q_1$, such that for all $\nu \in \{1, \ldots, m\}$, $i \in \mathcal{N}_{\nu}$, $z_{\nu i} \in \{0, 1\}$, $\bm{z}_{\mathcal{N}_{\nu i}} \in \{0,1\}^{d_{\nu i}},$  $$y_i(z_{\nu i}, \bm{z}_{\mathcal{N}_{\nu i}}) \ind Z_{\nu i}, \bm{Z}_{\mathcal{N}_{\nu i}} | \bm{X}_{\nu i}, \bm{b}_{\nu}^y.$$
\end{assumption}
Although inference in this work relies primarily upon the outcome model, we must also ensure appropriate covariate overlap, which tends to reduce the sensitivity of causal estimates to correct model specification \citep{li2023bayesian} and further ensures that the estimated treatment effect is not extrapolated to subregions of the covariate space for which there is little or no information. As the propensity score is central to the overlap or positivity assumption, we define the joint propensity score as $e(z_{\nu i}, \bm{z}_{\mathcal{N}_\nu i} | \bm{X}_{\nu i}, \bm{X}_{\mathcal{N}_{\nu i}}) = P(Z_{\nu i}=z_{\nu i}, \bm{Z}_{\mathcal{N}_{\nu i}} = \bm{z}_{\mathcal{N}_{\nu i}}| \bm{X}_{\nu i}, \bm{X}_{\mathcal{N}_{\nu i}})$, which corresponds to the probability of observing the joint treatment $(z_{\nu i}, \bm{z}_{\mathcal{N}_{\nu i}})$ conditional on individual and neighborhood covariates. Thus, we further require:
 \begin{assumption}[Positivity] For all $\nu \in \{1, \ldots, m\}, i \in \mathcal{N}$ and for all $z_{\nu i}, \bm{X}_{\nu i}, \bm{X}_{\mathcal{N}_{\nu i}}$, $$e(z_{\nu i}, \bm{z}_{\mathcal{N}_{\nu i}} | \bm{X}_{\nu i}, \bm{X}_{\mathcal{N}_{\nu i}}) > 0.$$
  \label{assum4}
\end{assumption}

 \section{Estimands}
 \label{section:estimands}
 To define the causal estimands of interest, we adopt the Bernoulli exposure allocation standardization \citep{liu2019doubly, lee2021estimating}. Let $\pi(\Sigma \bm{z}_{\mathcal{N}_{\nu i}}; \alpha) = \binom{d_{\nu i}}{\Sigma \bm{z}_{\mathcal{N}_{\nu i}}}\alpha^{\Sigma \bm{z}_{\mathcal{N}_{\nu i}}}(1-\alpha)^{d_{\nu i} -\Sigma \bm{z}_{\mathcal{N}_{\nu i}}}$ denote the probability of node $i$ in subgraph $\nu$ receiving neighborhood treatment $\Sigma \bm{z}_{\mathcal{N}_i}$ under Bernoulli allocation strategy or treatment coverage $\alpha$, which corresponds to the counterfactual probability of directly receiving treatment. Define the $i$-th individual's average potential outcome given individual exposure $z$ and allocation strategy $\alpha$ conditional on covariate value $\bm{x}$ as $$\mu_{z \alpha}^{\nu i} (\bm{x}) = \mathbb{E}\left[ \bar{y}_{\nu i}(z;\alpha) | \bm{X}_{\nu i} = \bm{x} \right] = \mathbb{E}  \left[ \sum_{\Sigma \bm{z}_{\mathcal{N}_{\nu i}} = 0 }^{d_{\nu i}} y_{\nu i}(z, \Sigma \bm{z}_{\mathcal{N}_{\nu i}}) \pi( \Sigma \bm{z}_{\mathcal{N}_{\nu i}}; \alpha) \biggr \rvert \bm{X}_{\nu i}= \bm{x} \right].$$
This second equality highlights the dependence of the potential outcomes on the individual covariates $\bm{x}$, that is, $y_{\nu i}(z, \Sigma \bm{z}_{\mathcal{N}_{\nu i}}) \equiv y_{\nu i}(z, \Sigma \bm{z}_{\mathcal{N}_{\nu i}}, \bm{x}_{\nu i})$, which will be made more explicit when specifying the generating model for the potential outcomes in Section \ref{section:inferentialmethod}. The average potential outcome in subgraph $\nu$ corresponds to the average of $\mu_{\nu i}^{z \alpha} (\bm{x})$ over the empirical subgraph covariate distribution $\hat{F}_X^{\nu}$, that is,  $\mu_{z \alpha}^{\nu} = \mathbb{E}\left[\mu_{z \alpha}^{\nu i} (\bm{x})\right] = \int \mu_{z \alpha}^{\nu i} (\bm{x}) \hat{F}_X^{\nu}(\bm{x}) d\bm{x} =  N_{\nu}^{-1} \sum_{i\in \mathcal{N}_{\nu}} \mu_{z \alpha}^{\nu i} (\bm{x}_{\nu i })$. The population average potential outcome is defined as $\mu_{z\alpha} = \mathbb{E}[\mu_{z \alpha}^{\nu}] = \int \mu_{z \alpha}^{\nu} \hat{H}(\nu) d\nu$, where $\hat{H}$ represents the empirical distribution of the subgraph-level average potential outcomes. This assigns equal weights to every subgraph-level potential outcome, such that $\mu_{z\alpha} = m^{-1} \sum_{\nu=1}^m \mu_{z \alpha}^{\nu}$. Note that $\mu_{z \alpha}^{\nu}$ and $\mu_{z \alpha}$ respectively correspond to subgraph-level and population-level average potential outcomes; \cite{li2023bayesian} refer to these as \emph{mixed} average potential outcomes to emphasize the fact that these models condition on the observed data (as is typical in the Bayesian framework) though we do marginalize over the empirical distribution of the observed  covariates $\bm{X}$, as can be seen in the expression for $\mu_{z \alpha}^{\nu}$. In this work, our primary interest lies in the inference for $\mu_{z\alpha}$ and marginal causal contrasts, rather than subgraph-specific estimands. We can also define the population-level marginal average potential outcome, which takes an additional expectation over the individual treatment allocation $Z$: $\mu_{\alpha} =  \alpha \mu_{1\alpha} + (1-\alpha) \mu_{0\alpha}$.

We consider various contrasts of the average potential outcomes to define the causal estimands \citep{hudgens2008toward, liu2019doubly, lee2021estimating, mcnealis2023doubly}. The direct exposure effect is defined as the difference between the average potential outcomes of untreated and treated individuals for a fixed allocation strategy $\alpha$, that is, $DE(\alpha) = \mu_{1\alpha} - \mu_{0\alpha}$. The spillover effect, also referred to as the \textit{indirect effect}, is defined as the difference between the average potential outcomes under counterfactual scenarios $\alpha$ and $\alpha'$ among the untreated, i.e., $IE(\alpha, \alpha') = \mu_{0\alpha}- \mu_{0\alpha'}$. Note that it would possible to define a spillover effect for the treated individuals as well. For allocation strategies $\alpha$ and $\alpha'$, we also define the total effect $TE(\alpha, \alpha') = \mu_{1\alpha}  - \mu_{0\alpha'}$ as well as the overall effect $OE(\alpha, \alpha') = \mu_{\alpha} - \mu_{\alpha'}$. As the name suggests, $TE(\alpha, \alpha')$ can be interpreted as the total effect of treatment, since $TE(\alpha, \alpha') = DE(\alpha) + IE(\alpha, \alpha')$, whereas the overall effect corresponds to the difference in average potential outcomes under one allocation strategy compared to another allocation strategy.

A summary of the different causal estimands presented along with the distributions being averaged over can be found in Table \ref{tab:notation} in Appendix \ref{appendix:notation}.
\section{Proposed joint mixed-effects model}
\label{section:inferentialmethod}

\begin{figure}[b]
\tikzset{
    -Latex,auto,node distance =1 cm and 1 cm,semithick,
    state/.style ={ellipse, draw, minimum width = 0.7 cm},
    point/.style = {circle, draw, inner sep=0.04cm,fill,node contents={}},
    bidirected/.style={Latex-Latex,dashed},
    el/.style = {inner sep=2pt, align=left, sloped}
}
\begin{tikzpicture}
    \node (1) at (0,0) {$Z_{\nu i}$};
    \node (2) at (0, -1) {$Z_{\nu j}$};
    \node (3) at (2, 0) {$Y_{\nu i}$};
    \node (4) at (2, -1) {$Y_{\nu j }$};
     \node[state] (5) at (-1.5, -0.5) {$\mathbf{b}_{\nu}^z$};
     \node[state] (6) at (3.5, -0.5) {$\mathbf{b}_{\nu}^y$};
    \path[-] (1) edge  (2);
    \path[-] (3) edge  (4);
    \path (2) edge (3);
    \path (1) edge (3);
    \path (2) edge (4);
    \path (1) edge (4);
    \path[dashed] (5) edge (1);
    \path[dashed] (5) edge (2);
    \path[dashed] (6) edge (3);
    \path[dashed] (6) edge (4);
     \path[bidirected] (5) edge[bend left=40] (6);
\end{tikzpicture}
\caption{Directed acyclic graph (DAG) for the joint linear mixed-effects model displaying interference and unmeasured heterogeneity in both treatment assignment
and outcome mechanisms within a subgraph. Baseline characteristics $\bm{X}_{\nu i}$ and $\bm{X}_{\nu j}$ are excluded from the figure for simplicity.}
\label{fig:dag}
\end{figure}
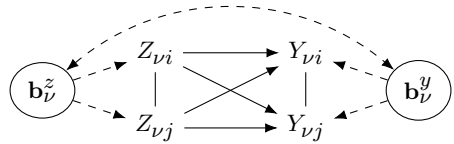

The proposed joint model allows for the possibility of unmeasured contextual confounding by assuming that the latent factors influencing the study outcome and the exposure selection in the social network are correlated as illustrated in the directed acyclic graph in Figure \ref{fig:dag}.
\subsection{Outcome model}
We assume that the potential outcomes arise according to
\begin{multline}
\label{eqn:outcome_model}
y_{\nu i}(z_{\nu i}, \Sigma \bm{z}_{\mathcal{N}_{\nu i}}) \\ = \beta_{Z}z_{\nu i} + h_1(\Sigma \bm{z}_{\mathcal{N}_{\nu i}}; \bm{\beta}_{\bm{Z}_{\mathcal{N}}}) + h_2(z_{\nu i} \Sigma \bm{z}_{\mathcal{N}_{\nu i}} ; \bm{\beta}_{Z \times \bm{Z}_{\mathcal{N}}}) + h_3(\bm{X}_{\nu i}^y; \bm{\beta}_{\bm{X}}) + \bm{W}_{\nu i}^{\top} \bm{b}_{\nu}^y + \delta_{\nu i},    
\end{multline}
where $h_1, h_2$ are known functions, $\bm{X}_{\nu i}^y$ is a vector of nodal covariates that are contained in $\bm{X}_{\nu i}$, $\bm{W}_{\nu i}$ is a vector of random effect covariates, and $\delta_{\nu i}$ is the random error of node $i$ in subgraph $\nu$. We further assume that $\bm{\delta} = (\bm{\delta}_{1}, \bm{\delta}_{2}, \ldots, \bm{\delta}_{m}) \sim N(\bm{0}, \bm{\Phi})$, where $\bm{\delta}_{\nu} = (\delta_{\nu 1}, \ldots, \delta_{\nu N_{\nu}})$, $\nu = 1, 2, \ldots, m$. In related work, the function $h_1$ is often assumed to be linear \citep{forastiere2021identification, lee2021estimating}, though non-linear functions such as spline representations could easily be accommodated. Similarly, $h_2$ and $h_3$ could either be linear or non-linear. According to this model, $\beta_Z$ and (potentially vector-valued) $\bm{\beta}_{\bm{Z}_{\mathcal{N}}}$ are tied to the direct and spillover effects of the exposure, respectively. The parameter $\bm{\beta}_{Z \times \bm{Z}_{\mathcal{N}}}$ captures the interaction between $z$ and $\Sigma \bm{z}_{\mathcal{N}_{\nu i}}$ and modifies the direct and interference effects with the level at which we fix the individual or neighborhood exposure. In the absence of interactions between the exposure variables and the covariates/random effects, the causal effects do not vary as a function of characteristics of the node or the subgraph.

We assume a simultaneous autoregressive (SAR) model for the vector of random errors $\bm{\delta}$, such that if $\bm{U} = \{ I_{N} - \tau \bm{A} \}^{-1}$ exists, the covariance matrix of $\bm{\delta}$ is given by \citep{cressie2015statistics} \begin{equation}
\label{eqn:SARcov}
    \bm{\Phi} = \sigma^2_{\varepsilon} \bm{U} \bm{U}^{\top}.
    \end{equation} 
    
    Due to its dependence on the adjacency matrix $\bm{A}$, the SAR covariance matrix is such that the potential outcomes of neighboring individuals are more correlated than those of non-neighbors \citep{yauck2021general}. According to this model, the parameter $\tau$ controls the strength of the interaction between neighbors \citep{f2007methods}. A simplifying assumption we make in simulations is that the subgraphs $G_{\nu}$ are disjoint, such that the sociomatrix $\bm{A}$ associated to $G$ is a block-diagonal matrix with $m$ blocks denoted $\bm{A}_{\nu}$, $\nu=1,\ldots m$, each corresponding to a subgraph $G_{\nu}$. This implies that $\bm{\Phi}$ is also block-diagonal with blocks $\bm{\Phi}_{\nu} = \sigma^2_{\varepsilon} \bm{U}_{\nu} \bm{U}_{\nu}^{\top}, \nu = 1, \ldots, m$, assuming that $\bm{U}_{\nu} = \{ I_{N_{\nu}} - \tau \bm{A}_{\nu}\}^{-1}$ exists for all $\nu$. If this assumption is not met, the computation might be infeasible in the presence of a large graph order $N$ due to difficulties with large-scale matrix inversion.

\subsection{Exposure model}
 We posit a Bernoulli distribution for the individual assigned treatment, such that $Z_{\nu i} | p_{\nu i } \sim \mathrm{Bernoulli}(p_{\nu i})$ with
\begin{equation}
\label{eqn:modelZ}
\mathrm{logit}(p_{\nu i}) = f(\bm{X}^z_{\nu i}; \bm{\gamma}) + \tilde{\bm{U}}_{\nu i}^{\top}\bm{b}_{\nu}^z,\end{equation}
where $p_{\nu i} = \mathbb{P}(Z_{\nu 
i } = 1 | \bm{X}^z_{\nu i}, \bm{b}_{\nu}^z) $, $f$ is a known function, and $\bm{U}_{\nu i}$ is a vector of random effect covariates. Note that the vector of covariates $\bm{X}^z_{\nu i}$ in the exposure selection model need not be identical to ${\bm{X}}_{\nu i}^y$. In a standard causal setting, it is typically preferable to avoid including instruments in the treatment model notably when implementing inverse weighting approaches, since it can lead to extreme weights and decreased precision in the estimation of causal effects \citep{kilpatrick2013exploring}. However, since we do not assume conditional exchangeability, a correctly specified model that follows the data-generating procedure (accounting for more than simply the confounders) for the treatment assignment mechanism is required for unbiased inference. Thus, $\bm{X}^z_{\nu i}$ is not necessarily a subset of ${\bm{X}}_{\nu i}^y$ in the proposed joint modeling methodology.
 \subsection{Random effect distribution}
The subgraph-level random effects of the above models, $\bm{b}_{\nu} = (\bm{b}_{\nu}^y, \bm{b}_{\nu}^z)$, are a $q_1 + q_2$ dimensional vector that capture the unobserved dependence among observations within each subgraph. We denote $\bm{b} = (\bm{b}_1, \ldots, \bm{b}_{m})$ the collection of subgraph-level random effects across the whole graph $G$. We assume that the random effects follow a multivariate normal distribution $\bm{b}_{\nu} \sim N_{q_1 + q_2}(\bm{0}, \Sigma)$, where
\begin{equation}
\label{eqn:randomeffectcov}
 \Sigma = \begin{pmatrix} \Sigma_{11} & \Sigma_{12} \\ \Sigma_{21} & \Sigma_{22} \end{pmatrix}.   
\end{equation}

The correlations of the random effects in the two models represent unobserved dependence among the outcome and the exposure assignment mechanisms. Groups defined by different treatment assignments are not, in the presence of such latent effects, conditionally exchangeable given observed covariates. This means that merely incorporating random effects in the outcome regression and ignoring the treatment assignment will not lead to valid inference of causal effects. Although based on Figure \ref{fig:dag}, conditioning on $\bm{b}^y$ would appear to block confounding, these are unobserved covariates and their posterior distribution may only be accurately estimated by accounting for the dependence between the outcomes and treatment assignments, the latter being informative for the imputation of missing potential outcomes \citep{follmann1995approximate}.  A model that estimates the random effects in the outcome without accounting for their correlation with the random effects of the treatment model will be poorly estimated, and this measurement error in the latent variables may lead to bias in the treatment effect estimators.

\subsection{Bayesian inference}
With interference, only one of multiple possible potential outcomes is observed for each node, with the number of potential outcomes for one unit depending on its degree by virtue of Assumption \ref{assum:stratified}. Under the Bayesian paradigm, unobserved potential outcomes are viewed as missing data, and inference about causal estimands is obtained from their posterior distributions \citep{rubin1978bayesian, ding2018causal, li2023bayesian, papadogeorgou2023spatial}. Let $\bm{y}(\cdot) = \{ y_i(\cdot) \text{ for all $i \in \mathcal{N}$}\}$ denote the collection of potential outcomes for all nodes in $G$ and possible individual and neighborhood exposures. We can then define the partition $\bm{y}(\cdot) = (\bm{Y}, \tilde{\bm{y}})$, where $\bm{Y}=(\bm{Y}_1, \bm{Y}_2, \ldots, \bm{Y}_{m})$ is the collection of observed outcomes across all the subgraphs and $\tilde{\bm{y}}$ is the collection of missing potential outcomes for all graph units. We denote the collection of covariate vectors for all the nodes in $G$ as $\bm{X}$. As we do not assign a probability distribution for the multi-dimensional $\bm{X}$, we consider a conditional likelihood in what follows. Following the specifications discussed in the previous section, the joint distribution of $(\bm{y}(\cdot), \bm{Z})$ conditional on $\bm{X}$, denoted $p(\bm{y}(\cdot), \bm{Z} | \bm{X}, \bm{\theta})$, is governed by the parameter vector $\bm{\theta} = (\bm{\beta}, \sigma^2_{\varepsilon}, \tau, \bm{\gamma}, \Sigma)$. Further, we can distinguish the parameters of the outcome model, $\bm{\theta}_Y=(\bm{\beta}, \sigma^2_{\varepsilon}, \tau, \Sigma)$, and those of the exposure model, $\bm{\theta}_Z=(\bm{\gamma}, \Sigma)$, where $\Sigma$ is shared between the two models. Bayesian causal inference proceeds by specifying a prior $p(\bm{\theta})$ and imputing missing potential outcomes from the posterior predictive distribution
\begin{align*}
p(\tilde{\bm{y}} | \bm{O}) 
&=  \int_{\bm{\theta}} p(\tilde{\bm{y}}\ |\ \bm{\theta}, \bm{O}) p(\bm{\theta} | \bm{O}) d\bm{\theta}\\
& \propto \int_{\bm{\theta}} \left[\int_{\bm{b}_{\nu}} p(\tilde{\bm{y}} |  \bm{b},  \bm{\theta}, \bm{O}) p(\bm{b}_{\nu} | \bm{\theta}, \bm{O}) d\bm{b}\right] \mathcal{L}(\bm{\theta} | \bm{O}) p(\bm{\theta}) d\bm{\theta},
\end{align*}
where $p(\tilde{\bm{y}}\ |\ \bm{\theta}, \bm{O})$ is the sampling distribution of the missing potential outcomes given the parameters and the observed data, $p(\bm{\theta} | \bm{O})$ is the posterior distribution of $\bm{\theta}$, $\mathcal{L}(\bm{\theta} | \bm{O}) = p(\bm{O} | \bm{\theta})$ is the observed data likelihood, and $\bm{O} = (\bm{Y}, \bm{Z}, \bm{X})$ is the observed data. Since our interest lies in marginal causal estimands, we perform inference from the marginal posterior predictive distribution of the potential outcome with respect to the random effects, which is highlighted by the proportionality statement. 

Typically, the central component in Bayesian inference for causal effects is the specification of the outcome model \citep{li2023bayesian}. In instances where the propensity score model is ignorable, unbiased inference can be obtained about causal inference without having to specify a model for the propensity of treatment. However, the presence of dependent latent effects among the outcome and exposure models threatens the inference, since the exposure assignment mechanism is informative for the imputation of missing potential outcomes. To see this, we turn to the observed data likelihood and write
\begin{align*}
\label{jmm}
\mathcal{L}(\bm{\theta} | \bm{O})  &=\prod_{\nu = 1}^m \int_{\bm{b}_{\nu}} p(\bm{Y}_{\nu}, \bm{Z}_{\nu} | \bm{X}_{\nu}, \bm{b}_{\nu}, \bm{\theta}) p(\bm{b}_{\nu} | \bm{\theta}) d\bm{b}_{\nu} \\ &= \prod_{\nu}^m \int_{\bm{b}_{\nu}} \prod_{i=1}^{N_{\nu}} p(Y_{\nu i} | Z_{\nu i} = z_{\nu_i}, \Sigma \bm{Z}_{\mathcal{N}_{\nu i}} = \Sigma \bm{z}_{\mathcal{N}_{\nu i}},  \bm{X}_{\nu i}^y, \bm{b}_{\nu}^y, \bm{\theta}) p(Z_{\nu i} | \bm{X}^z_{\nu i}, \bm{b}_{\nu}^z, \bm{\theta})  p(\bm{b}_{\nu} | \bm{\theta}) d\bm{b}_{\nu}\\
&= \prod_{\nu}^m \int_{\bm{b}_{\nu}} \prod_{i=1}^{N_{\nu}} p(y_{\nu i}(z_{\nu_i},  \Sigma \bm{z}_{\mathcal{N}_{\nu i}}) |   \bm{X}_{\nu i}^y, \bm{b}_{\nu}^y, \bm{\theta}_Y) p(Z_{\nu i} | \bm{X}^z_{\nu i}, \bm{b}_{\nu}^z, \bm{\theta}_Z)  p(\bm{b}_{\nu} | \Sigma) d\bm{b}_{\nu},
\end{align*}
where the third equality follows from counterfactual consistency, latent exchangeability, and positivity. This means that having access to the latent confounders $\bm{b}$ would render the treatment assignment ignorable. However, since $\bm{b}$ is unobserved and yet a component of the treatment assignment mechanism through $\Sigma$, the treatment assignment is informative for the imputation of missing potential outcomes and needs to be accounted for in the Bayesian procedure \citep{papadogeorgou2023spatial, xu2023causal, ricciardi2020bayesian, mccandless2007bayesian}. Hence, in the presence of unobserved contextual confounding, the propensity score model is \textit{not} ignorable in the inference for $\mu_{z\alpha}$, $\mu_{\alpha}$ and functions of these quantities.

For node $i$ in subgraph $\nu$, $\tilde{y}_{\nu i}(z_{\nu i}, \Sigma \bm{z}_{\mathcal{N}_{\nu i}})$ acts like a missing variable that can be imputed after setting exposure to some value $(z_{\nu i}, \Sigma \bm{z}_{\mathcal{N}_{\nu i}})$ and sampling from the marginal sampling distribution for the replicated potential outcomes given parameters, assigned treatments and observed outcomes
\begin{align*}
    p(\tilde{y}_{\nu i}(z_{\nu i}, \Sigma \bm{z}_{\mathcal{N}_{\nu i}}) | \bm{\theta},\bm{O}) 
    =  \int_{\bm{b}_{\nu}} p(\tilde{y}_{\nu i} | Z_{\nu i} = z_{\nu i}, \Sigma \bm{Z}_{\mathcal{N}_{\nu i}} = \Sigma \bm{z}_{\mathcal{N}_{\nu i}}, \bm{b}_{\nu},  \bm{\theta}, \bm{O}) p(\bm{b}_{\nu} | \bm{\theta}, \bm{O}) d\bm{b},
    \end{align*}    
where the equality holds under latent ignorability because of the inclusion of the propensity score model in the posterior inference for $\bm{\theta}$.

\subsection{Prior specification}
We adopt a combination of vague and weakly informative priors for model parameters. These parameters are the coefficients of the outcome model in (\ref{eqn:outcome_model}), the coefficients of the exposure model in (\ref{eqn:modelZ}), the residual variance and the autocorrelation parameters of the SAR covariance matrix in (\ref{eqn:SARcov}), and the parameters of the random effect covariance matrix in (\ref{eqn:randomeffectcov}). 

For the coefficients of the outcome and exposure models, $\bm{\beta}$ and $\bm{\gamma}$, we adopt vague independent $N(0, \sigma^2_{\text{prior}})$ prior distributions with $\sigma^2_{\text{prior}}=100$. We specify $\tau \sim \mathrm{Uniform}[-1,1]$ and $\sigma_{\varepsilon} \sim \text{Half-Cauchy}(\alpha_{\text{prior}})$ with the recommended hyperparameter of $\alpha_{\text{prior}}=25$, which results in a reasonably flat prior distribution \citep{gelman2006prior, polson2012half}. 

For the covariance matrix of the random effects, we propose using the decomposition $\Sigma = SRS$, where $S$ is a $(q_1 + q_2) \times (q_1 + q_2)$ diagonal matrix with the standard deviations of the random effects as its diagonal elements and $R$ is a correlation matrix, and assigning the prior $R \sim \text{LKJ}(\eta_{\text{prior}})$ with $\eta_{\text{prior}}=1$, where LKJ refers to the Lewandowski-Kurowicka-Joe distribution \citep{lewandowski2009generating}. This choice of prior assigns equal density to all correlation matrices of order $q_1 + q_2$. Finally, we assign independent $\text{Half-Cauchy}(\alpha_{\text{prior}})$ priors to the diagonal elements of $S$. In the simulations and application that follow, we consider the joint random intercept model, i.e., $\bm{b}_{\nu} = (b_{\nu}^y, b_{\nu}^z)$,  such that $$ S= \begin{pmatrix} \sigma_{b^y} & 0 \\ 0 & \sigma_{b^z} \end{pmatrix}, \quad R = \begin{pmatrix} 1 & \ \rho  \\ \rho \ & 1 \end{pmatrix}.$$ In that case, the LKJ prior reduces to assigning a $\mathrm{Uniform}[-1, 1]$ prior to $\rho$.

The resultant posterior distributions are approximated through computational approaches. We employ Markov chain Monte Carlo (MCMC) methods. 
\
\subsection{Bayesian standardization}
\label{subsection:bayesstandardization}
In this section, we describe a procedure to recover the posterior distributions of $\mu_{z \alpha}$, $\mu_{\alpha}$, and relevant constrasts from the posterior distribution of $\bm{\theta}$. In a point treatment setting, the Bayesian g-formula or standardization typically entails marginalizing the individual average potential outcome over the distribution of covariates, which requires obtaining the posterior distributions of $\bm{\theta}$ and $\bm{\theta}_X$. However, recall that in this work, we do not model the multi-dimensional covariates $\bm{X}$ and instead rely upon mixed average causal effects, which are reasonable approximations to population average causal effects \citep{li2023bayesian}.

Samples from the posterior distributions of $\mu_{z \alpha}$, $\mu_{\alpha}$, and relevant contrasts are obtained by marginalizing the posterior distribution of $\mu_{z\alpha}(\bm{x}) \equiv \mu_{z\alpha}(\bm{x}; \bm{\theta})$ over the empirical distribution of $\bm{X}$ and the subgraph distribution, where the notation highlights the dependence on the outcome model parameter. The standardization procedure essentially consists of computing the following integral 
$$\tilde{\mu}_{z\alpha} = \int_{\nu} \int_{\bm{X}} \mu_{z \alpha}^{\nu i} (\bm{x}; \bm{\theta}_Y) \hat{F}_X^{\nu}(\bm{x}) \hat{H}(\nu) d\bm{x} d\nu = m^{-1} \sum_{\nu = 1}^m N_{\nu}^{-1} \sum_{i =1}^{N_{\nu}} \tilde{\mu}_{z \alpha}^{\nu i} (\bm{x}_{\nu i}; \bm{\theta}),$$
where $\tilde{\mu}_{z \alpha}^{\nu i} (\bm{x}_{\nu i}; \bm{\theta}) =  \sum_{\Sigma \bm{z}_{\mathcal{N}_{\nu i}} = 0 }^{d_{\nu i}} \tilde{y}_{\nu i}(z, \Sigma \bm{z}_{\mathcal{N}_{\nu i}}) \pi( \Sigma \bm{z}_{\mathcal{N}_{\nu i}}; \alpha)$ is the imputed individual average potential outcome and $\tilde{y}_{\nu i}(z, \Sigma \bm{z}_{\mathcal{N}_{\nu i}}) \equiv \tilde{y}_{\nu i}(z, \Sigma \bm{z}_{\mathcal{N}_{\nu i}}, \bm{x}_{\nu i})$ is a random draw from the marginal sampling distribution for the replicated potential outcomes given individual treatment $z$, neighborhood treatment $\Sigma \bm{z}_{\mathcal{N}_{\nu i}}$, and covariate vector $\bm{x}_{\nu i}$. We use Monte Carlo integration to obtain samples from the marginal distribution for the potential outcomes with respect to the random effects. The Bayesian standardization procedure is summarized as follows.

\begin{enumerate}
\item Set an exposure value $z$ and an allocation strategy $\alpha$.
\item For each posterior draw $\bm{\theta}^{(s)}$, $s=1,\ldots, S$, perform the following steps:
\begin{enumerate}
\item Sample $\tilde{\bm{\delta}}^{(s)} \sim  N(\bm{0}_{N}, \bm{\Phi}^{(s)})$ with $\bm{\Phi}^{(s)}= \sigma^2_{\varepsilon} \bm{U}_{s} \bm{U}_{s}^{\top}$ and $\bm{U}^{(s)} = \{ I_{N} - \tau^{(s)} \bm{A}\}^{-1}$.
\item For each subgraph $\nu \in \{1, \ldots m\}$, sample a random effect vector $\tilde{\bm{b}}^{b  {(s)}}_{\nu} \sim N(\bm{0}, \Sigma^{(s)})$. 
\item For each subgraph $\nu \in \{1, \ldots m\}$, each node $i \in \{1, \ldots, N_{\nu}\}$, and each neighborhood exposure value $\Sigma \bm{z}_{\mathcal{N}_{\nu i}} \in \{0, \ldots, d_{\nu i}\}$, draw the corresponding missing potential outcome from the conditional posterior predictive distribution 
\begin{multline*}\tilde{y}_{\nu i}^{b{(s)}}(z, \Sigma \bm{z}_{\mathcal{N}_{\nu i}}) \\ = \beta_{Z}^{(s)}z + h_1(\Sigma \bm{z}_{\mathcal{N}_{\nu i}}; \bm{\beta}_{\bm{Z}_{\mathcal{N}}}^{(s)}) + h_2(z \Sigma \bm{z}_{\mathcal{N}_{\nu i}} ; \bm{\beta}_{Z \times \bm{Z}_{\mathcal{N}}}^{(s)}) + h_3(\bm{x}_{\nu i}; \bm{\beta}_{\bm{X}}^{(s)}) + \bm{W}_{\nu i}^{\top} \tilde{\bm{b}}_{\nu}^{y, b{(s)}}+ \tilde{\delta}_{\nu i}^{(s)}.
\end{multline*}
\item Repeat steps b) and c) $B$ times for each exposure value $(z, \Sigma \bm{z}_{\mathcal{N}_{\nu i}})$ to obtain a draw from the sampling distribution of the missing potential outcomes conditional on $\bm{\theta}$ (marginalized with respect to $\bm{b}$) for all nodes and all subgraphs using
$$ \tilde{y}_{\nu i}^{(s)}(z, \Sigma \bm{z}_{\mathcal{N}_{\nu i}}) = \frac{1}{B}\sum_{b=1}^B \tilde{y}_{\nu i}^{b{(s)}}(z, \Sigma \bm{z}_{\mathcal{N}_{\nu i}}).$$
\end{enumerate}
\item Using the imputed potential outcomes in Step 2, for each posterior draw $s = 1, \ldots, S$, subgraph $\nu \in \{1,\ldots, m\}$, and node $i \in \{1,\ldots, N_{\nu}\}$, obtain the posterior individual average potential outcome as
$$ \tilde{\mu}_{z \alpha}^{\nu i {(s)}} (\bm{x}_{\nu i}) = \sum_{\Sigma \bm{z}_{\mathcal{N}_{\nu i}} = 0 }^{d_{\nu i}} \tilde{y}_{\nu i}^{(s)}(z, \Sigma \bm{z}_{\mathcal{N}_{\nu i}}) \pi( \Sigma \bm{z}_{\mathcal{N}_{\nu i}}; \alpha).$$
\item Finally, obtain the posterior population average potential outcome for treatment $z$ and coverage $\alpha$ as
$\tilde{\mu}_{z\alpha}^{(s)} = m^{-1} \sum_{\nu = 1}^m N_{\nu}^{-1} \sum_{i =1}^{N_{\nu}} \tilde{\mu}_{z \alpha}^{\nu i {(s)}} (\bm{x}_{\nu i}),$
as well as the posterior population marginal average potential outcome, $\tilde{\mu}_{\alpha}^{(s)}= \alpha \tilde{\mu}_{1\alpha}^{(s)} + (1-\alpha) \tilde{\mu}_{0\alpha}^{(s)}$.
\end{enumerate}
Approximations of the posterior distributions for relevant causal contrasts can be obtained by applying relevant functions ($DE(\alpha)$, $IE(\alpha, \alpha')$,  $TE(\alpha, \alpha')$, $OE(\alpha, \alpha')$) to each posterior draw $\tilde{\mu}_{z\alpha}^{(s)}$ and $\tilde{\mu}_{\alpha}^{(s)}$, $s=1,\ldots, S$. 

\section{Simulations}
\label{section:simulations}
We present simulation results under the joint random intercept model, where $\bm{b}_{\nu} = (b_{1\nu}, b_{2 \nu})$ to assess frequentist properties of the proposed Bayesian estimator based on the joint mixed-effects model (JMM) in finite samples compared to estimators based on linear mixed- and fixed effects models (LMM, FEM, respectively). The default simulation is conducted according to the following steps with $m=50$ disjoint subgraphs, such that $\bm{A}$ is block-diagonal. 
\begin{enumerate}
\item \textbf{Generation of the network}: For $\nu \in \{1,\ldots, 30\}$, we sampled $N_{\nu}$ from the distribution $\mathrm{Poisson}(35)$, and for $\nu \in \{31,\ldots, 50\}$, we sampled $N_{\nu}$ from the distribution $\mathrm{Poisson}(12)$. For each subgraph, we generated the random vector $\bm{H}_{\nu}$ of size $N_{\nu}$ whose elements are from a $\mathrm{Bernoulli}(0.5)$ distribution. The $m$ subgraphs were generated according to an exponential-family random graph model in which matching $\bm{H}$ values increase the probability of tie between two nodes, inducing network homophily. The final network $G$ was taken as the union of the $m$ components. 
\item For $\nu =1,\ldots, m$, we sampled $(b^y_{\nu}, b^z_{\nu})$ from $N_2(\bm{0}, \Sigma)$, with $$\Sigma = \begin{pmatrix} \sigma^2_{b^y} &  \sigma_{b^y} \sigma_{b^z} \rho \\ \sigma_{b^y} \sigma_{b^z} \rho &  \sigma^2_{b^z}\end{pmatrix}$$
and sampled the SAR errors $\bm{\delta}_{\nu}$ from $N_{N_{\nu}}(\bm{0}, \sigma^2_{\varepsilon}\bm{U}_{\nu}\bm{U}_{\nu}^{\top})$.
\item For $\nu =1,\ldots, m, \ i = 1, \ldots, N_{\nu}$, two baseline covariates were generated as $X_{1\nu i} \sim \mathcal{N}(0,1)$ and $X_{2\nu i} \sim \mathrm{Bernoulli}(0.5)$. Define the proportion of treated nodes in a neighborhood as $p(\bm{z}_{\mathcal{N}_{\nu i}}) = \Sigma \bm{z}_{\mathcal{N}_{\nu i}}/d_{\nu i}$.  Potential outcomes for all $z_{\nu i} \in \{0,1\}$, $\Sigma \bm{z}_{\mathcal{N}_{\nu i}} \in \{0,1,\ldots, d_{\nu i}\}$ were given by
$$y_{\nu i}(z_{\nu i}, \bm{z}_{\mathcal{N}_{\nu i}}) = 2 + 2 z_{\nu i} + p(\bm{z}_{\mathcal{N}_{\nu i}}) + z_{\nu i} p(\bm{z}_{\mathcal{N}_{\nu i}}) -1.5 |X_{1i}| + 2X_{2i} - 3 |X_{1i}|X_{2i} + b_{\nu}^y + \delta_{\nu i},$$
where $\delta_{\nu i}$ is the $i$-th entry of the SAR errors vector $\bm{\delta}_{\nu}$.
\item  The treatment was generated as $Z_{\nu i} = \mathrm{Bernoulli}(p_{\nu i})$, where $$p_{\nu i} = \mathrm{logit}^{-1}\left[0.1 +0.2 |X_{1i}| + 0.2X_{2i}|X_{1i}| - H_{\nu i}+ b_{\nu}^{z} \right]. $$ Under this model, nodes that share ties tended to have similar treatment assignments.
\item Based on the potential outcomes defined in Step 2, observed outcomes were set equal to $Y_{\nu i} = y_{\nu i}\left(Z_{\nu i},  \Sigma \bm{Z}_{\nu i}\right)$.
\end{enumerate}

In the standard simulation scenario, simulations were carried out 500 times by repeating steps 2-4 with $\mathrm{Var}(b_{\nu}^y)=\mathrm{Var}(b_{\nu}^z)=1$ and a correlation coefficient $\rho=0, 0.2, 0.5, 0.8$ between the outcome and exposure model random intercepts. For simplicity, we first set conditionally independent random errors for observations within each component with $\tau = 0$ and $\sigma^2_{\varepsilon} = 1$, such that the covariance matrix of the random error vector within a component was equal to $\sigma^2_{\varepsilon} \bm{U}_{\nu} \bm{U}_{\nu}^{\top} = I_{N_{\nu}}$. The true parameters were obtained by averaging the potential outcomes that we generated in Step 2 over the simulated data sets. For each simulated data set and each model (JMM, LMM, and FEM), we fit four chains with 10,000 iterations used as burn-in period, and thinning by 50 iterations. The posterior distributions of $\mu_{z\alpha}$ and $\mu_{\alpha}$ for $z=0,1$ and $\alpha \in \{0.1, 0.2, \ldots, 0.9\}$ as well as relevant causal contrasts were obtained from the JMM, LMM, and FEM parameter posterior distributions, respectively, using the procedure detailed in Subsection \ref{subsection:bayesstandardization} with $B=50$ Monte Carlo samples. For all estimands of interest, the Bayesian estimator was defined as the posterior mean and we calculated the posterior standard deviation and a 95\% equi-tailed credible interval as measures of uncertainty.

Table \ref{tab:tab1} shows the relative bias (RB, \%), the average of posterior standard deviations (ASD), the empirical standard deviation of Bayesian estimators (ESD), and the coverage of 95\% credible intervals (ECP, \%) associated with JMM, LMM, and FEM for the estimation of $\mu_{0,0.7}$, $\mu_{1,0.7}$, and $DE(0.7) = \mu_{1,0.7} - \mu_{0,0.7}$. Our discussion of results focuses on the direct effect $DE(0.7)$. In general, JMM provides valid inference in comparison to LMM or FEM and as expected, the benefit associated with accounting for contextual confounding becomes clearer as the correlation $\rho$ increases. The relative bias associated with LMM, albeit never very large, increases with $\rho$, while the relative bias associated with JMM remains small. In general, the posterior standard deviations accurately estimate the variability of the regression estimators when either JMM or LMM is used, whereas the actual variability of the causal estimators is drastically underestimated when the multilevel structure of the data is not accounted for (FEM). 

\begin{figure}[b]
\includegraphics[scale=0.46]{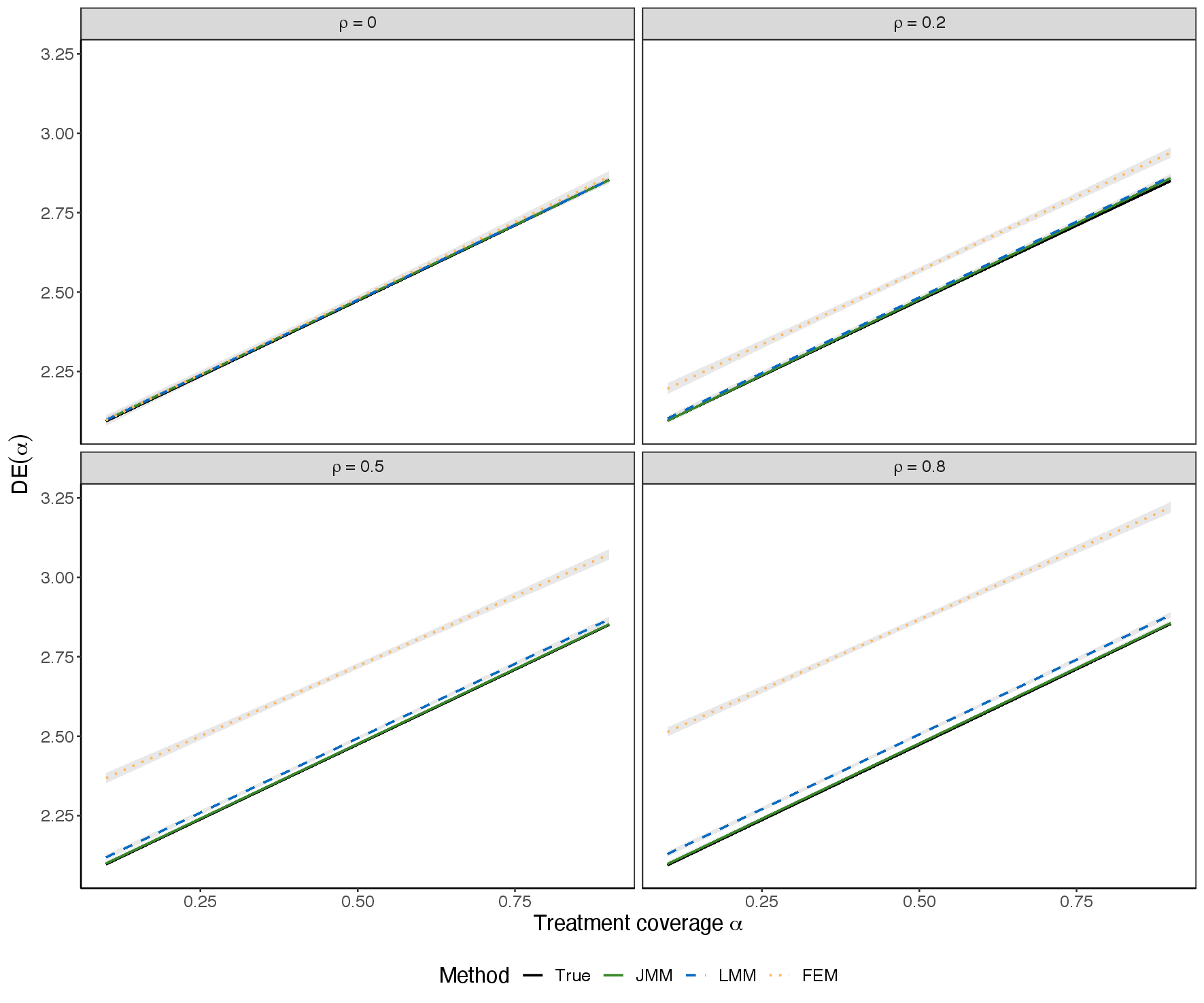}
\caption{Average JMM, LMM, and FEM estimates of the dose response curve $DE(\alpha)$ for $\rho = 0, 0.2, 0.5, 0.8$. The black solid line represents the true effect. The light gray area represents 95\% pointwise frequentist confidence bands based on $S= 500$ simulation replicates. Note that the bands are very narrow in the case of JMM and LMM. Also, notice the near overlap of JMM with the true dose-response curve in all four scenarios.}
\label{fig:de}
\end{figure}

\begin{figure}[t]
\includegraphics[scale=0.46]{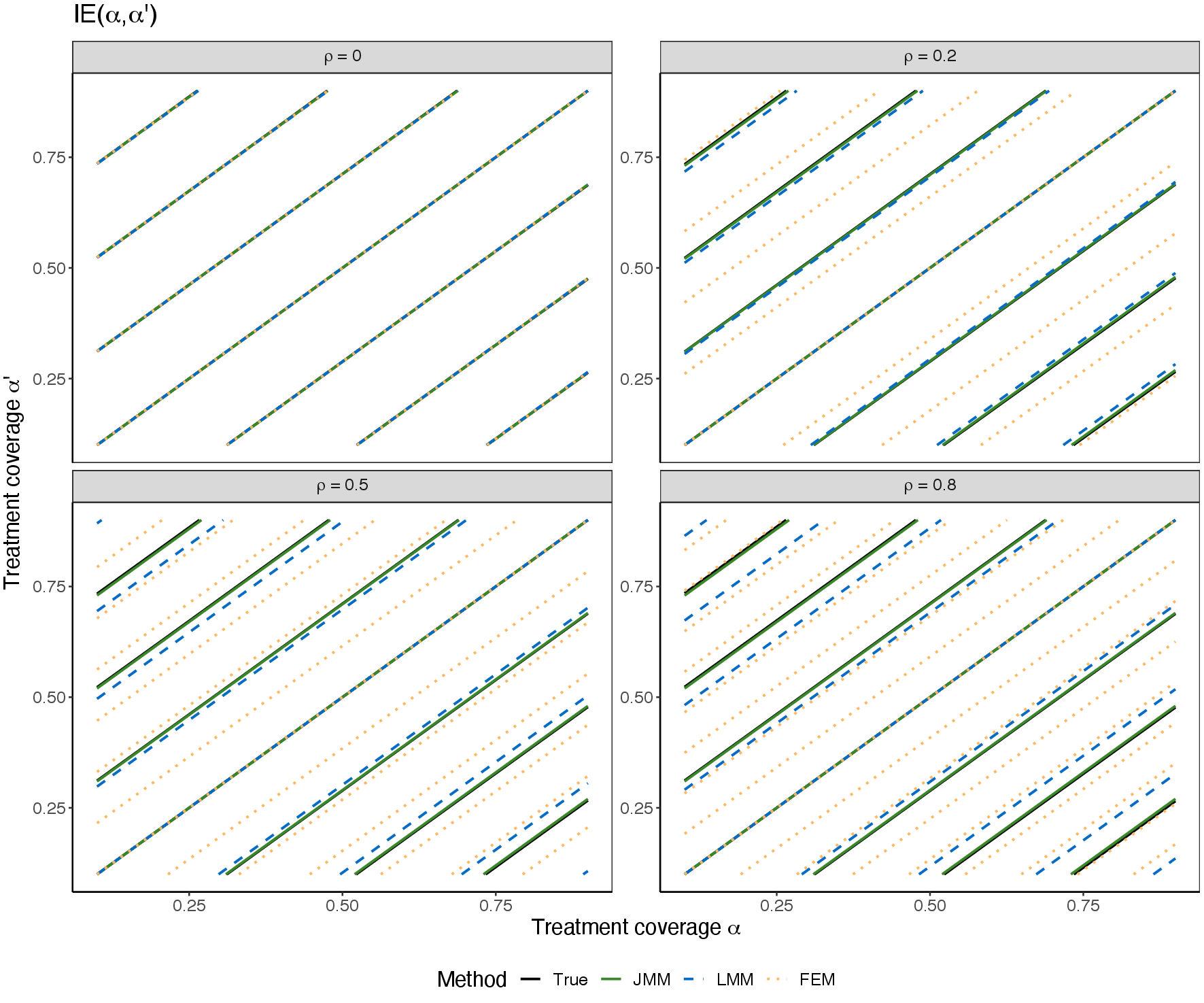}
\caption{Average JMM, LMM, and FEM estimates of the surface $IE(\alpha, \alpha')$ for $\rho = 0, 0.2, 0.5, 0.8$. The black solid line represents the true effect. Notice the near overlap of JMM with the true surface in all four scenarios.}
\label{fig:ie}
\end{figure}

Figure \ref{fig:de} displays the average of the 500 JMM, LMM, and FEM estimators of the direct effect $DE(\alpha)$ as well as the corresponding true dose-response curve, where each panel corresponds to one of the $\rho$ values considered. Figure \ref{fig:ie} displays the average of the 500 JMM, LMM, and FEM estimators of the indirect effect $IE(\alpha, \alpha')$ as well as the corresponding true surface. LMM and FEM estimators of the direct effect sustain a positive bias as soon as there is presence of unmeasured contextual confounding ($\rho > 0$) and this bias increases as the contextual confounding intensifies. The same conclusions pertain to the indirect effect, which takes the form of a surface as a function of $\alpha$ and $\alpha'$.

In addition to the simulation scenario above, we generated data sets under different a variety of conditions with a fixed $\rho=0.2$. We considered the following six scenarios for which results can be found in Table \ref{tab:tab1}:
\begin{enumerate}
\item We assumed a SAR structure for the random errors with $\tau = 0.1$ and modeled them as such across the different methods (JMM, LMM, and FEM). Comparing this scenario to the standard scenario, JMM and LMM's performance remain largely similar. 
\item We reduced the number of subgraphs to $m=30$. For $\nu \in \{1,\ldots, 20\}$, we sampled $N_{\nu}$ from the distribution $\mathrm{Poisson}(35)$, and for $\nu \in \{21,\ldots, 30\}$, we sampled $N_{\nu}$ from the distribution $\mathrm{Poisson}(12)$. With fewer clusters, the ASD and ESD of the estimators show a slight increase. 
\item We generated data sets using a larger random effect variance with $\sigma^2_{b^y}=\sigma^2_{b^z} = 4$, which also caused the ASD and ESD of the estimators to increase. Note the worsened performance of FEM in this setting.
\item In an additional scenario, we generated data under no treatment effect with the following outcome model
$$y_{\nu i}(z_{\nu i}, \bm{z}_{\mathcal{N}_{\nu i}}) = 2 -1.5 |X_{1i}| + 2X_{2i} - 3 |X_{1i}|X_{2i} + b_{\nu}^y + \delta_{\nu i}.$$
However, we included terms for the treatment, the neighborhood treatment and the interaction at the modeling stage. The average Bayesian estimate of $DE(0.7)$ is closest to 0 with JMM, followed by LMM and FEM. 
\item In another scenario, $(b_{\nu}^z, b_{\nu}^y)$ were simulated from a bivariate exponential distribution with scale parameters $\lambda_{b^y}=\lambda_{b^z}=1$ for the marginal distributions and a correlation of 0.2 using the R package \texttt{MDBED} \citep{mdbedpackage}. The random effects were then centered to have mean 0. At the modeling stage, we used the regular method based on a multivariate normal likelihood to investigate the impact of misspecifying the random effect distribution. The performance of JMM, LMM, and FEM appears to be largely unaffected by the misspecification of the random effect distribution for the set simulation parameters.
\item Lastly, to mimic certain aspects of the Add Health study, we also generated the potential outcomes under a truncated normal distribution with conditional mean
$$2 + 2 z_{\nu i} + p(\bm{z}_{\mathcal{N}_{\nu i}}) + z_{\nu i} p(\bm{z}_{\mathcal{N}_{\nu i}}) -1.5 |X_{1i}| + 2X_{2i} - 3 |X_{1i}|X_{2i} + b_{\nu}^y,$$
standard deviation of $1$, a lower truncation bound of $-5$ and an upper truncation bound of $9$. Again, we obtain comparable results as the main simulation scenario under this simulation scheme.
\end{enumerate}
As observed across a wide range of scenarios and different forms of model misspecification, JMM outperforms the competitor approaches of LMM and FEM whenever contextual confounding is present.

\begin{table}[t]
\centering
\caption{Finite sample properties of the Bayesian estimator based on correctly specified joint mixed-effects model (JMM), linear  model (LMM), and fixed effects model (FEM) for 500 simulations. }
\scriptsize
\begin{tabular}{c c c c c c c c c c c c c c }
\toprule
$\rho$ & Method & \multicolumn{4}{c}{$\mu_{0, 0.7}$} & \multicolumn{4}{c}{$\mu_{1, 0.7}$} &  \multicolumn{4}{c}{$DE(0.7)$}   \\
    \cmidrule(l){3-6} \cmidrule(l){7-10} \cmidrule(l){11-14}
& &  RB & ASD & ESD & ECP   &  RB & ASD & ESD & ECP  &  RB & ASD & ESD & ECP   \\
\midrule 
 0 &  \multicolumn{5}{l}{Main scenario}  \\ 
 & JMM & -0.32 & 0.16 & 0.17 & 92.8 & -0.06 & 0.16 & 0.17 & 94.6 & 0.07 & 0.08 & 0.08 & 96.2 \\ 
 &LMM& -0.21 & 0.16 & 0.17 & 93.2 & -0.02 & 0.16 & 0.17 & 94.6 & 0.08 & 0.08 & 0.08 & 96.6 \\ 
 &FEM& -0.46 & 0.09 & 0.19 & 64.8 & 0.11 & 0.08 & 0.21 & 53.6 & 0.39 & 0.10 & 0.13 & 85.8 \\ 
   \midrule
   
0.2   & \multicolumn{5}{l}{Main scenario}  \\
 & JMM & -0.35 & 0.16 & 0.16 & 95.2 & 0.04 & 0.16 & 0.16 & 94.8 & 0.23 & 0.08 & 0.07 & 97.4 \\ 
 &LMM& -0.05 & 0.16 & 0.16 & 94.4 & 0.29 & 0.16 & 0.16 & 94.2 & 0.45 & 0.08 & 0.07 & 97.4 \\ 
 &FEM& 1.83 & 0.09 & 0.19 & 62.6 & 2.93 & 0.08 & 0.21 & 49.6 & 3.46 & 0.10 & 0.13 & 77.8 \\ 
    & \multicolumn{5}{l}{SAR errors ($\tau = 0.1$)}  \\
  & JMM  & -0.81 & 0.20 & 0.16 & 97.5 & -0.24 & 0.20 & 0.16 & 98.0 & 0.04 & 0.08 & 0.08 & 95.5 \\ 
 & LMM& -0.78 & 0.20 & 0.16 & 97.5 & -0.00 & 0.20 & 0.16 & 98.0 & 0.37 & 0.08 & 0.08 & 94.0 \\ 
 &FEM & -0.37 & 0.08 & 0.17 & 63.0 & 1.28 & 0.08 & 0.17 & 58.0 & 2.08 & 0.08 & 0.10 & 87.0 \\ 
   & \multicolumn{5}{l}{Lower sample size ($m=30$)}  \\
 & JMM &  -0.29 & 0.22 & 0.22 & 94.4 & -0.08 & 0.22 & 0.23 & 93.8 & 0.03 & 0.10 & 0.09 & 97.6 \\ 
 &LMM & 0.01 & 0.21 & 0.22 & 94.2 & 0.16 & 0.21 & 0.23 & 93.4 & 0.24 & 0.10 & 0.09 & 97.6 \\ 
&FEM&  1.58 & 0.11 & 0.25 & 58.4 & 2.87 & 0.10 & 0.27 & 49.6 & 3.49 & 0.12 & 0.16 & 81.0 \\ 
   & \multicolumn{5}{l}{Larger random effect variance ($\sigma^2_{b^y}=\sigma^2_{b^z} = 4$)}  \\
   &JMM & -0.17 & 0.30 & 0.30 & 95.0 & -0.09 & 0.30 & 0.30 & 94.8 & -0.06 & 0.10 & 0.09 & 97.6 \\ 
  &LMM &0.35 & 0.30 & 0.30 & 95.0 & 0.25 & 0.30 & 0.30 & 94.0 & 0.20 & 0.10 & 0.09 & 97.6 \\ 
  &FEM & 2.50 & 0.15 & 0.39 & 53.0 & 6.27 & 0.11 & 0.38 & 35.0 & 8.07 & 0.16 & 0.29 & 60.8 \\ 
   & \multicolumn{5}{l}{No treatment effect\textsuperscript{b} }  \\
&JMM& -0.34 & 0.16 & 0.17 & 93.8 & 0.23 & 0.16 & 0.17 & 93.0 & - & 0.08 & 0.08 & 95.0 \\ 
&LMM&  -0.18 & 0.16 & 0.17 & 94.4 & 1.44 & 0.16 & 0.17 & 91.8 & - & 0.08 & 0.08 & 94.8 \\ 
&FEM&  2.78 & 0.09 & 0.19 & 64.8 & 19.71 & 0.08 & 0.20 & 47.4 & - & 0.10 & 0.14 & 70.6 \\ 
   & \multicolumn{5}{l}{Bivariate exponential random effect distribution}  \\
  &JMM& -0.35 & 0.15 & 0.08 & 99.8 & -0.05 & 0.15 & 0.09 & 100.0 & 0.10 & 0.09 & 0.08 & 97.8 \\ 
 &LMM& -0.64 & 0.15 & 0.08 & 99.6 & 0.09 & 0.15 & 0.09 & 100.0 & 0.43 & 0.09 & 0.08 & 97.6 \\ 
 &FEM& -4.42 & 0.09 & 0.12 & 80.8 & 0.85 & 0.09 & 0.12 & 80.4 & 3.35 & 0.11 & 0.14 & 79.6 \\ 
   & \multicolumn{5}{l}{Truncated normal outcome distribution}  \\
  &JMM & -0.20 & 0.15 & 0.15 & 96.2 & -0.09 & 0.15 & 0.15 & 95.8 & -0.03 & 0.08 & 0.07 & 96.4 \\ 
 &LMM& -0.09 & 0.15 & 0.15 & 95.6 & 0.12 & 0.15 & 0.15 & 96.0 & 0.22 & 0.08 & 0.07 & 96.6 \\ 
&FEM&  1.49 & 0.08 & 0.17 & 61.0 & 2.68 & 0.07 & 0.18 & 48.0 & 3.28 & 0.10 & 0.13 & 82.2 \\ 
   \midrule 
0.5  &  \multicolumn{5}{l}{Main scenario}  \\ 
& JMM & 0.13 & 0.16 & 0.17 & 92.8 & 0.10 & 0.16 & 0.17 & 93.4 & 0.09 & 0.08 & 0.08 & 96.0 \\ 
 &LMM& 0.41 & 0.16 & 0.17 & 93.0 & 0.61 & 0.16 & 0.16 & 93.2 & 0.71 & 0.08 & 0.08 & 95.0 \\ 
 &FEM& 4.98 & 0.09 & 0.18 & 62.2 & 7.57 & 0.08 & 0.20 & 20.4 & 8.81 & 0.10 & 0.13 & 35.2 \\ 
   \midrule
   0.8  &  \multicolumn{5}{l}{Main scenario}  \\
   &JMM&  -0.59 & 0.16 & 0.17 & 93.0 & -0.08 & 0.16 & 0.17 & 93.2 & 0.16 & 0.08 & 0.08 & 95.4 \\ 
 &LMM& 0.11 & 0.16 & 0.16 & 93.0 & 0.83 & 0.16 & 0.16 & 92.6 & 1.17 & 0.08 & 0.08 & 93.8 \\ 
&FEM&  7.21 & 0.08 & 0.16 & 61.2 & 12.02 & 0.07 & 0.18 & 3.0 & 14.32 & 0.09 & 0.14 & 7.2 \\   
  \bottomrule
\end{tabular}
\label{tab:tab1}
\begin{tablenotes}
    \item  \textsuperscript{a} For the main simulation scenario, the true estimand values were $\mu_{0,0.7} = 1.27$, $\mu_{1,0.7} = 3.93$, $DE(0.7)=2.66$.
    \item \textsuperscript{b} For this scenario, the small value for the true direct effect ($DE(0.7) = 0.00051$) precluded the meaningful calculation of a relative bias. The average Bayesian estimates of $DE(0.7)$ were $0.00397$, $0.010431$, and $0.10381$ for JMM, LMM, and FEM, respectively.
  \end{tablenotes}
\end{table}

\section{Evaluation of spillover effects of maternal education in Add Health}
\label{section:application}
We apply the estimators proposed in Section \ref{section:inferentialmethod} to estimate the causal effects of college maternal education on academic performance. We investigate this question while aiming to mitigate bias from potentially missing subgraph-level confounders. In the following, the outcome is a student's GPA and the exposure is defined as the binary indicator of whether their mother has completed four years of college. The students who did not report
living with their biological mother, stepmother, foster mother, or adoptive mother at the time of the survey were excluded from the analytic sample. We also excluded isolates from the sample for computational reasons. There was partial actor non-response in the data set, which occurs when some, but not all, attribute information is available for one or more nodes in the network \citep{krause2018missing}. To remedy this, a single imputation of the nodal attributes was performed using imputation by chained equations with predictive mean matching and 20 cycles. The second column of Table \ref{tab:table12} provides descriptive statistics for the nodal attributes after the imputation. The network structure of the Add Health subsamble of $m=16$ schools had 29 connected components with 7,931 students and 26,384 shared connections (average degree was 6.65) after the exclusions. Among the 7,931 students, 2,873 (36.2\%) had a college-educated mother. Figure \ref{fig:prop_treated} displays a histogram of the observed proportion of friends with a college-educated in the analytical sample, which shows a right-skewed distribution. The following baseline covariates were included in the outcome model: sex, age, student's race (as a proxy for the race of the mother), whether the student was adopted, whether the mother was born in the US, whether the father is at home, household size, whether the mother works for pay, student's perception of how much their mother cares for them, club participation, motivation at school, trouble at school, sense of belonging, attendance, health status, physical fitness, and screen time. The model for the exposure only included four of the baseline covariates: student's race (as a proxy for the race of the mother), whether the mother was born in the US, whether the father is at home, and household size.

We considered three analyses (JMM, LMM, and FEM), all of which adjust for the same baseline covariates enumerated above and include all variables linearly in the predictors of the outcome model (and of the propensity score model in the case of JMM). The main analyses do not impose a SAR structure on the random errors, since the adjacency matrix in the Add Health subnetwork is not block-diagonal (there are ties between students from different schools) and its inversion would be prohibitively computationally expensive considering its large dimension ($N=7,931$). 

\begin{figure}[b]
\includegraphics[scale=0.30]{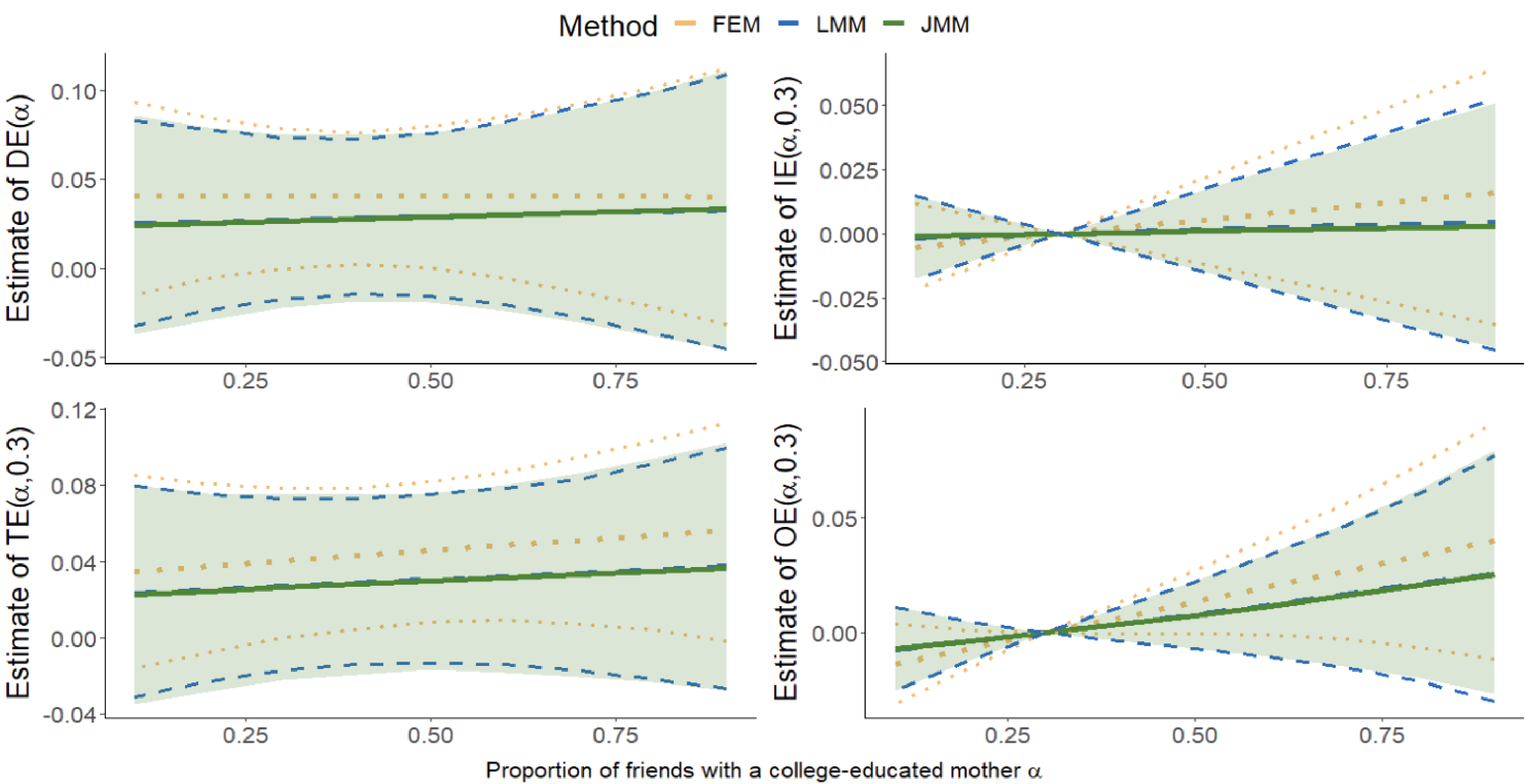}
\caption[JMM, LMM, and FEM estimates of ${DE}(\alpha)$, ${IE}(\alpha, 0.3)$, ${TE}(\alpha, 0.3)$, and ${OE}(\alpha, 0.3)$ effects with 95\% credible intervals.]{JMM, LMM, and FEM estimates of ${DE}(\alpha)$, ${IE}(\alpha, 0.3)$, ${TE}(\alpha, 0.3)$, and ${OE}(\alpha, 0.3)$ effects with 95\% credible intervals.
        }
\label{fig:addhealthresults}
\end{figure}

Estimates based on the JMM, LMM, and FEM approaches for ${DE}(\alpha)$, ${IE}(\alpha, 0.3)$, ${TE}(\alpha, 0.3)$ and ${OE}(\alpha, 0.3)$ along with 95\% credible intervals are shown in Figure \ref{fig:addhealthresults} and Table \ref{tab:estimatesAddHealth}. In general, JMM, LMM, and FEM estimators give similar conclusions, in that the directions of the dose-response curves align. The point estimates suggest that academic performance is increased not only directly by having a college-educated mother, but also indirectly by having a greater proportion of friends with a college-educated mother. We provide an interpretation for $DE(0.2)$. If every student in this network were to be ``assigned'' a college-educated mother versus a mother with a non-college education, we would expect the average GPA difference to be 0.0252 (95\% CrI: [-0.0291, 0.0792]) points under a counterfactual probability of receiving treatment $\alpha=20$\%. Using the JMM method, we obtain an estimate of 0.0025 (95\% CrI: [-0.0373, 0.0428]) for $IE(0.8, 0.3)$. This implies that if we held the maternal education indicator to 0, the GPA of students with 80\% of friends with college-educated mothers would be on average higher by 0.0025 points compared to a setting in which the maternal education indicator setting is again set to 0 with however 30\% of friends having college-educated mothers, assuming both groups' covariate distribution matched that of the overall population.

Most credible intervals shown in Table \ref{tab:estimatesAddHealth} contain zero, with the exception of $DE(0.4)$ and $TE(\alpha, 0.3)$ for $\alpha = 0.4, 0.6, 0.8$ when FEM is used. In general, effect estimates are closer to zero based on JMM and LMM compared to the FEM approach, which ignores the clustering in the graph. Effect estimates based on our approach (JMM) are also lesser in magnitude than those based on LMM, which assumes ignorability of the exposure assignment. In the JMM approach, the posterior median for $\rho$ was $0.58$ (95\% CrI: $[0.09, 0.84]$). These results suggest that there might exist unmeasured contextual confounding, and our approach partially mitigates the bias due to the unmeasured latent traits. Also, the credible intervals for $IE(\alpha, 0.3)$ based on JMM are slightly narrower than those for LMM, illustrating that accounting for unmeasured contextual confounding can also lead to efficiency gains in the estimation of interference effects.

\begin{table}{t}
\caption{\label{tab:estimatesAddHealth} JMM, LMM, and FEM estimates for the Add Health study along with 95\% credible intervals for ${DE}(\alpha)$, ${IE}(\alpha, 0.3)$, ${TE}(\alpha, 0.3)$ and ${OE}(\alpha, 0.3)$.}
\centering
\begin{tabular}{lccc}
  \toprule
 Estimand & JMM & LMM & FEM \\ 
  \midrule
  $DE(\alpha)$\\
$\alpha = 0.2$ & 0.0252 (-0.0291, 0.0792) & 0.0260 (-0.0233, 0.0778) & 0.0405 (-0.0056, 0.0848) \\ 
   $\alpha = 0.4$ & 0.0277 (-0.0187, 0.0748) & 0.0280 (-0.0143, 0.0729) & 0.0405 (0.0020, 0.0764) \\ 
 $\alpha = 0.6$ & 0.0302 (-0.0243, 0.0813) & 0.0300 (-0.0207, 0.0820) & 0.0405 (-0.0055, 0.0851) \\ 
  $\alpha = 0.8$ & 0.0326 (-0.0380, 0.0993) & 0.0319 (-0.0362, 0.0989) & 0.0404 (-0.0218, 0.1020) \\ 
 \midrule
 $IE(\alpha, 0.3)$\\
$\alpha = 0.2$ & -0.0005 (-0.0086, 0.0074) & -0.0008 (-0.0089, 0.0074) & -0.0027 (-0.0108, 0.0059) \\ 
 $\alpha = 0.4$ & 0.0005 (-0.0074, 0.0085) & 0.0008 (-0.0074, 0.0087) & 0.0027 (-0.0058, 0.0109) \\ 
 $\alpha =0.6$ & 0.0015 (-0.0223, 0.0257) & 0.0023 (-0.0226, 0.0262) & 0.0080 (-0.0177, 0.0325) \\ 
   $\alpha =0.8$ & 0.0025 (-0.0373, 0.0428) & 0.0038 (-0.0379, 0.0439) & 0.0133 (-0.0295, 0.0542) \\ 
    \midrule
     $TE(\alpha, 0.3)$\\
     $\alpha = 0.2$ & 0.0247 (-0.0281, 0.0760) & 0.0252 (-0.0228, 0.0755) & 0.0378 (-0.0077, 0.0807) \\ 
  $\alpha = 0.4$ & 0.0282 (-0.0190, 0.0752) & 0.0288 (-0.0136, 0.0728) & 0.0432 (0.0043, 0.0784) \\ 
  $\alpha= 0.6$ & 0.0317 (-0.0182, 0.0804) & 0.0323 (-0.0139, 0.0787) & 0.0485 (0.0096, 0.0870) \\ 
  $\alpha= 0.8$ & 0.0351 (-0.0230, 0.0941) & 0.0357 (-0.0216, 0.0915) & 0.0537 (0.0043, 0.1035) \\ 
   \midrule
   $OE(\alpha, 0.3)$\\
$\alpha = 0.2$ & -0.0034 (-0.0118, 0.0050) & -0.0037 (-0.0118, 0.0046) & -0.0068 (-0.0147, 0.0011) \\ 
 $\alpha = 0.4$ & 0.0037 (-0.0037, 0.0113) & 0.0039 (-0.0035, 0.0112) & 0.0067 (-0.0002, 0.0137) \\ 
  $\alpha =0.6$ & 0.0117 (-0.0101, 0.0344) & 0.0122 (-0.0103, 0.0340) & 0.0202 (-0.0009, 0.0413) \\ 
  $\alpha =0.8$ & 0.0207 (-0.0196, 0.0625) & 0.0213 (-0.0209, 0.0609) & 0.0335 (-0.0061, 0.0729) \\ 
   \bottomrule
\end{tabular}
\end{table}
\section{Discussion}
\label{section:discussion}
In this paper, we developed a Bayesian approach for causal inference with clustered network-based observational studies, estimating the direct and indirect effects of a binary exposure on a
continuous outcome while accounting for potential unmeasured contextual confounding.  We discussed the inherent challenges that arise in causal inference with multilevel network data, which are prevalent in education, and discussed the relevance of accounting for potential nonignorability of the treatment assignment. We proposed to simultaneously model the outcome and treatment assignment by JMM, which reduces unmeasured contextual confounding when models are correctly specified. Our proposed methodology can also address potential network autocorrelation in the outcome. We described a Bayesian standardization procedure which can be used to obtain the posterior distributions of causal estimands of interest from the posterior distributions of the JMM parameters. We note that JMM has a specific representation of unmeasured contextual confounding, the level of which is governed by the level of heterogeneity in the outcome and treatment assignment mechanisms and the correlation between the treatment assignment and the outcome. In the case of the joint random intercept model, this correlation  operates through the correlation parameter $\rho$. Large random effect variances or $\rho$ values may lead to significant bias due to unmeasured confounding. The simulation study demonstrated that JMM provides valid inference in a wide variety of settings for average potential outcomes and causal contrasts, where LMM and FEM may not. A particular advantage of our proposed approach is that it performs well with either the presence or absence of unmeasured contextual confounding.

Our proposal fits more widely into the literature of Bayesian causal inference methods incorporating a model for the propensity score. Existing ways of combining propensity score and outcomes models comprise including the propensity score as a covariate in the outcome model, specifying dependent priors or shared parameters between the propensity score and outcome models, or posterior predictive inference through inverse probability weighting and doubly robust estimation \citep{saarela2015bayesian, saarela2016bayesian, li2023bayesian}. As we used a latent Gaussian structure to relate the outcome and propensity score models, our method falls under the second category of methods \citep{xu2023causal}. By allowing the information to flow from the outcome model to the estimation the propensity score model, we open the door to potential \textit{model feedback}, a problem specific to the Bayesian causal inference setting \citep{zigler2013model, li2023bayesian}. Feedback can introduce bias in causal effect estimators when the outcome is misspecified. To remedy this, two-stage procedures have been proposed and entail fitting a Bayesian model for the propensity score and plugging posterior draws of the propensity score in the outcome \citep{zigler2014uncertainty, li2023bayesian, stephens2023causal}. However, it is unclear whether a pure joint modeling approach can be avoided in the context of violation of strong ignorability of the treatment assignment.


As previously mentioned, JMM relies on parametric assumptions regarding the joint distribution of outcomes and treatment assignments. Unbiased estimation depends on correct specification of the outcome data-generating process, the treatment assignment mechanism, and the random-effects distribution. Under the assumption of strong ignorability, parametric model assumptions would not necessarily be required for valid causal inference. However, in the context of a school-based design such as Add Health, unmeasured school-level confounders are highly likely as students attending the same school may share latent characteristics affecting both their outcome and exposure. To relax the unconfoundedness assumption, the tradeoff here is to make additional assumptions on model specifications. As it is more restrictive and also computationally expensive, the gain provided by JMM may not be worthwhile in larger networks or when contextual confounding is weak. Another caveat specific to our analysis of spillover effects of maternal education in Add Health is that we had to use a smaller subset of the data set for computational reasons, potentially leading to an underpowered analysis.

Despite the potential merits of this work, methodological challenges remain in order to improve inference for causal effects in the presence of network interference. Of the different sources of network confounding \citep{vanderweele2013social}, our approach only addresses contextual confounding. Another challenge inherent to this context is homophily bias. Homophily is defined as the increased tendency of units with similar characteristics of forming ties. In sociological settings, this can lead to latent variable dependence due to latent traits that both drive the formation of network ties and predict an individual's treatment or outcome. In all scenarios we considered here, we assumed that all variables driving homophily were measured and conditioned upon in analyses, which is unlikely to be the case in practice. Imposing a SAR structure on the random errors of the outcome model might account for some of the latent dependence, but more work has to be done to better understand how homophily can affect inference in these settings. Latent homophily poses a threat to the identification of causal effects in various ways. \cite{shalizi2011homophily} showed that unless all of the nodal
attributes driving both social-tie formation and the behavior of interest are observed, then peer effects are generally unidentified. Assuming that the social network is generated by a latent community, \cite{mcfowland2023estimating} proposed asymptotically unbiased estimators of peer influence under homophily by controlling the latent location of
each node. Another avenue for future work would be to use their methodology and incorporate estimates of latent positions in the outcome and exposure models to control for potential latent homophily when inferring for interference effects.


\section*{Acknowledgements}

This research was enabled in part by support provided by Calcul Québec (\url{https://www.calculquebec.ca/}) and the Digital Research Alliance of Canada (\url{https://alliancecan.ca/en}). The case study in this paper uses data from Add Health, funded by grant P01 HD31921 (Harris) from the Eunice Kennedy Shriver National Institute of Child Health and Human Development (NICHD), with cooperative funding from 23 other federal agencies and foundations. Add Health is currently directed by Robert A. Hummer and funded by the National Institute on Aging cooperative agreements U01 AG071448 (Hummer) and U01AG071450 (Aiello and Hummer) at the University of North Carolina at Chapel Hill. Add Health was designed by J. Richard Udry, Peter S. Bearman, and Kathleen Mullan Harris at the University of North Carolina at Chapel Hill.\\

\bibliographystyle{imsart-nameyear} 
\bibliography{bibliography}       

\newpage
\setcounter{table}{0}
\setcounter{figure}{0}
\setcounter{equation}{0}
\renewcommand{\thefigure}{S\arabic{figure}}
\renewcommand{\thetable}{S\arabic{table}}
\renewcommand{\theequation}{S\arabic{equation}}
\begin{appendix}
\section{Summary of notation used in the causal framework}
\label{appendix:notation}
\begin{table}[H]
\centering
\begin{threeparttable}
\caption{Breakdown of the different causal estimands and the distributions being averaged over}
\begin{tabular}{llll}
\toprule
Estimand\tnote{a } & Notation & Definition & Distributions being averaged over\tnote{b }\\
\midrule
Individual APO under $z$ and $\alpha$ & $\mu_{z \alpha}^{\nu i} (\bm{x})$ & $\mathbb{E}\left[ \bar{y}_{\nu i}(z;\alpha) | \bm{X}_{\nu i} = \bm{x} \right]$ &  Not applicable\\
Subgraph-level APO under $z$ and $\alpha$ & $\mu_{z\alpha}^{\nu}$ & $\int \mu_{z \alpha}^{\nu i} (\bm{x}) \hat{F}_X^{\nu}(\bm{x}) d\bm{x}$ & $\hat{F}_X^{\nu}$ \\
Population-level APO under $z$ and $\alpha$ & $\mu_{z\alpha}$ & $\int \mu_{z \alpha}^{\nu} \hat{H}(\nu) d\nu$ & $\hat{F}_X^{\nu}$, $\hat{H}$ \\
Marginal population-level APO under $\alpha$ & $\mu_{\alpha}$ & $\alpha \mu_{1 \alpha} + (1-\alpha) \mu_{0 \alpha}$  & $\hat{F}_X^{\nu}$, $\hat{H}$, and $F_{Z, \alpha}$ \\
Direct effect & $DE(\alpha)$& $ \mu_{1 \alpha} - \mu_{0 \alpha}$ &  $\hat{F}_X^{\nu}$, $\hat{H}$ \\
Indirect effect & $IE(\alpha)$ & $ \mu_{0 \alpha} - \mu_{0 \alpha'}$ & $\hat{F}_X^{\nu}$, $\hat{H}$ \\
Total effect & $TE(\alpha)$ &  $\mu_{1 \alpha} - \mu_{0 \alpha'}$ & $\hat{F}_X^{\nu}$, $\hat{H}$ \\ 
Overall effect & $OE(\alpha)$ & $  \mu_{\alpha} - \mu_{ \alpha'}$ & $\hat{F}_X^{\nu}$, $\hat{H}$, and $F_{Z, \alpha}$ \\ 
\bottomrule
\end{tabular}
\begin{tablenotes}
\item \textsuperscript{a} APO: Average potential outcome; $z$ and $\alpha$ correspond to the counterfactual treatment and the treatment coverage, respectively, where the treatment coverage represents the counterfactual probability of directly receiving the treatment.
\item \textsuperscript{b} $\hat{F}_X^{\nu}$: Empirical covariate distribution in subgraph $\nu$; $\hat{H}$: Empirical distribution of the subgraph-level potential outcomes; $F_{Z, \alpha}$: Distribution of the counterfactual treatment assignment under treatment coverage $\alpha$, namely the Bernoulli distribution with parameter $\alpha$. 
\end{tablenotes}
    \label{tab:notation}
\end{threeparttable}
\end{table}

\section{Descriptives for the Add Health analytical data set}
\label{section:appendixa}
\begin{table}[H]
\centering
\begin{threeparttable}
\caption{Characterization of the Add Health subnetwork of 7,931 students after isolates}
    \small
    \begin{tabular}{l l c c }
    \toprule
       Global network  & Schools & \multicolumn{2}{c}{16} \\
      characteristics & Nodes & \multicolumn{2}{c}{7,931} \\
       & Components & \multicolumn{2}{c}{29}\\
         & Edges & \multicolumn{2}{c}{26,384}
\\
         & Average Degree (SD) & \multicolumn{2}{c}{6.65 (3.88)}\\
         & Edge density & \multicolumn{2}{c}{8.39$\cdot 10^{-4}$}\\
         & Transitivity & \multicolumn{2}{c}{0.171} \\
         & Assortativity\tnote{a}& \multicolumn{2}{c}{0.123} \\
         \midrule
         Between-school  & Nodes (Median, Q1-Q3) & \multicolumn{2}{c}{382.0 (234.5 - 602.5)}\\
         network characteristics & Edges (Median, Q1-Q3) & \multicolumn{2}{c}{1221.50 (814.75 - 1859.25)}\\
         & Average Degree (Mean, SD) & \multicolumn{2}{c}{6.56 (0.99)} \\
         & Edge density (Median, Q1-Q3) &  \multicolumn{2}{c}{0.015 (0.010 - 0.031)}\\
         & Transitivity (Median, Q1-Q3) &  \multicolumn{2}{c}{0.204 (0.157 - 0.241)}\\
         & Assortativity\tnote{a} $\ $(Median, Q1-Q3) & \multicolumn{2}{c}{0.061 (0.013 - 0.080)}\\
         \bottomrule
    \end{tabular}
\begin{tablenotes}
\item \textsuperscript{a} The assortativity coefficient is based on maternal education.
\end{tablenotes}
    \label{tab:table1}
\end{threeparttable}
\end{table}

\begin{table}[H]
\centering
\begin{threeparttable}
\caption{Descriptive statistics of nodal attributes in the Add Health subnetwork of 7,931 students}
    \scriptsize
    \begin{tabular}{l l c c }
    \toprule
         & & Before imputation & After single imputation\\
         \cmidrule(l){3-3} \cmidrule(l){4-4}
                  Grade Point Average  & Mean (SD) &  2.94 (0.76) & 2.92 (0.76)\\
                  \cmidrule(l){2-2}  \cmidrule(l){3-3} \cmidrule(l){4-4}
         & Valid responses & 7,219 (91.0\%) & 7,931 (100.0\%)\\
         & Missing & 712 (9.0\%) 
 & 0 (0.0\%) \\
                  \midrule
         Maternal education & 4-year college degree &  2,607 (36.8\%)  & 2,873 (36.2\%)  \\
         & High school graduate or less &  4,478 (63.2\%) 
 & 5,058 (63.8\%)\\
  \cmidrule(l){2-2}  \cmidrule(l){3-3} \cmidrule(l){4-4}
         & Valid responses & 7,085 (89.3\%) & 7,931 (100.0\%)\\
         & Missing & 846 (10.7\%) & 0 (0.0\%) \\
         \midrule
         Age  & Mean (SD) &  14.80 (1.64)
 & 14.81 (1.64)\\
                  \cmidrule(l){2-2}  \cmidrule(l){3-3} \cmidrule(l){4-4}
         & Valid responses & 7,912 (99.8\%) & 7,931 (100.0\%)\\
         & Missing & 19 (0.2\%) 
 & 0 (0.0\%) \\
  \midrule
         Sex & Male &  3,857 (48.9\%) 
 &   3,878 (48.9\%) 

  \\
         & Female &   4,033 (51.1\%) 
 
 &  4,053 (51.1\%) 
\\
  \cmidrule(l){2-2}  \cmidrule(l){3-3} \cmidrule(l){4-4}
         & Valid responses & 7,890 (99.5\%) & 7,931 (100.0\%)\\
         & Missing & 41 (0.5\%) & 0 (0.0\%) \\
                  \midrule
         Race & White & 3,956 (49.9\%) & 3,956 (49.9\%) \\
         & Black &   2,399 (30.2\%) & 2,399 (30.2\%)  \\
         & Asian &    282 (3.6\%)  & 282 (3.6\%) \\
         & Other &   1,294 (16.3\%)
 & 1,294 (16.3\%) \\
   \cmidrule(l){2-2}  \cmidrule(l){3-3} \cmidrule(l){4-4}
 & Valid responses & 7,931 (100.0\%)& 7,931 (100.0\%)\\
         & Missing & 0 (0.0\%)& 0 (0.0\%) \\
         \midrule
         Lives with father & Yes &  6,003 (76.3\%)  & 6,042 (76.2\%) \\
         & No &  1,868 (23.7\%)  & 1,889 (23.8\%)\\
          \cmidrule(l){2-2}  \cmidrule(l){3-3} \cmidrule(l){4-4}
 & Valid responses & 7,871 (92.2\%)& 7,931  (100.0\%)\\
         & Missing & 60 (0.8\%)& 0 (0.0\%) \\
         \midrule 
         Mother was born in the US & Yes &  7,039 (90.0\%) & 7,128 (89.9\%) \\
         & No &  786 (10.0\%)  & 803 (10.1\%) \\
         \cmidrule(l){2-2}  \cmidrule(l){3-3} \cmidrule(l){4-4}
 & Valid responses & 7,825 (98.7\%)& 7,931  (100.0\%)\\
         & Missing & 106 (1.3\%)& 0 (0.0\%) \\
         \bottomrule
    \end{tabular}
    \label{tab:table12}
\end{threeparttable}
\end{table}

\begin{figure}[h]
\includegraphics[scale=0.4]{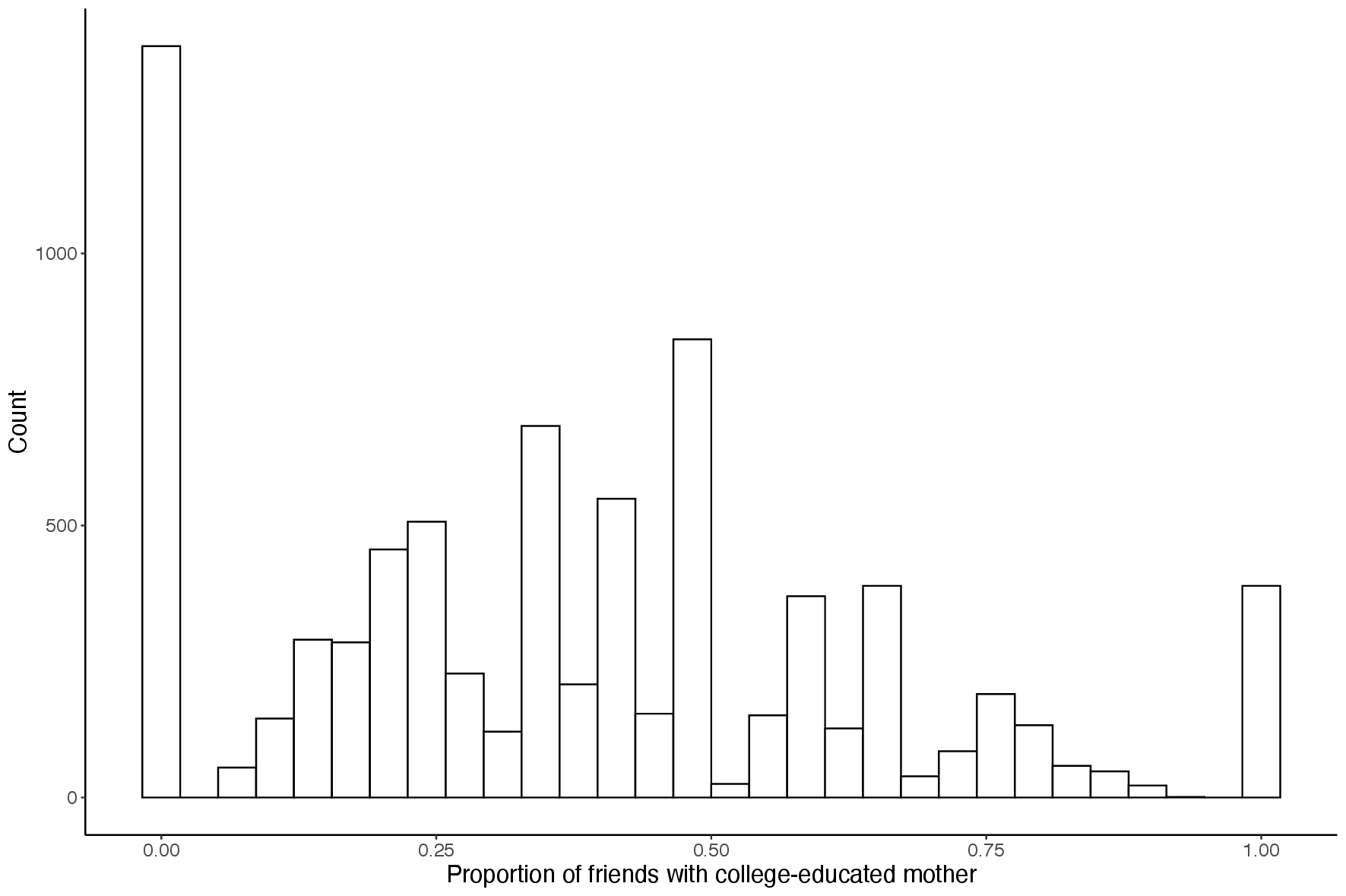}
\caption{Distribution of the observed proportion of friends with a college-educated mother in the Add Health subnetwork.
        }
\label{fig:prop_treated}
\end{figure}

\end{appendix}

\end{document}